\documentclass[utf8]{frontiersSCNS} % for Science, Engineering and Humanities and Social Sciences articles

\setcitestyle{square} % for Physics and Applied Mathematics and Statistics articles
\usepackage{url,microtype,subcaption}
\usepackage[onehalfspacing]{setspace}

\def\keyFont{\fontsize{8}{11}\helveticabold }
\def\firstAuthorLast{Pudritz and Ray}
\def\Authors{Ralph E Pudritz\,$^{1,*}$ and Tom P Ray\,$^{2}$}
\def\Address{$^{1}$Department of Physics and Astronomy, McMaster University, Hamilton, Ontario, Canada \\
$^{2}$School of Cosmic Physics, Dublin Institute for Advanced Studies, Ireland}

\def\corrAuthor{Corresponding Author}

\def\corrEmail{pudritz@mcmaster.ca}

\newcommand{\pc}{{\rm\thinspace pc}}

\newcommand{\au}{{\rm ~ au}}

\begin{document}
\onecolumn
\firstpage{1}

\title[Magnetic Fields in Outflows]{The Role of Magnetic Fields in Protostellar Outflows and Star Formation} 

\author[\firstAuthorLast ]{\Authors} %This field will be automatically populated
\address{} %This field will be automatically populated
\correspondance{} %This field will be automatically populated

\extraAuth{}% If there are more than 1 corresponding author, comment this line and uncomment the next one.
%\extraAuth{corresponding Author2 \\ Laboratory X2, Institute X2, Department X2, Organization X2, Street X2, City X2 , State XX2 (only USA, Canada and Australia), Zip Code2, X2 Country X2, email2@uni2.edu}

\maketitle

\begin{abstract}

%%% Leave the Abstract empty if your article does not require one, please see the Summary Table for full details.
%\section{}
\noindent 
The role of outflows in the formation of stars and the protostellar disks that generate them is a central question in astrophysics.  Outflows are associated with star formation across the entire stellar mass spectrum. In this review, we describe the observational, theoretical, and computational advances on magnetized outflows, and their role in the formation of disks and stars of all masses in turbulent, magnetized clouds.  The ability of torques exerted on disks by magnetized winds to efficiently extract and transport disk angular momentum was developed in early theoretical models and confirmed by a variety of numerical simulations.  The recent high resolution Atacama Large Millimeter Array (ALMA) observations of disks and outflows now confirm several key aspects of these ideas, e.g. that jets rotate and originate from large regions of their underlying disks. New insights on accretion disk physics show that magneto-rotational instability (MRI) turbulence is strongly damped, leaving magnetized disk winds as the dominant mechanism for transporting disk angular momentum. This has major consequences for star formation, as well as planet formation.  Outflows also play an important role in feedback processes particularly in the birth of low mass stars and cluster formation. Despite being almost certainly fundamental to their production and focusing, magnetic fields in outflows in protostellar systems, and even in the disks, are notoriously difficult to measure.  Most methods are indirect and lack precision, as for example, when using optical/near-infrared line ratios. Moreover, in those rare cases where direct measurements are possible -  where synchrotron radiation is observed, one has to be very careful in interpreting derived values. Here we also explore what is known about magnetic fields from observations, and take a forward look to the time when facilities such as SPIRou and the SKA are in routine operation.

%For full guidelines regarding your manuscript please refer to \href{http://www.frontiersin.org/about/AuthorGuidelines}{Author Guidelines}.

\tiny
 \keyFont{ \section{Keywords:} outflows, protostellar jets, magnetic fields, star formation, disks, disk winds, feedback, planet formation} %All article types: you may provide up to 8 keywords; at least 5 are mandatory.
\end{abstract}

\section{Introduction}
 
Of the many important roles that magnetic fields play in the formation of stars, perhaps none is more dramatic nor as full of consequence as is the launch and collimation of powerful outflows.  It is now nearly 70 years since the first evidence for energetic outflows in star formation regions was discovered \citep{1951ApJ...113..697H,1952ApJ...115..572H} although not realized as such. In fact it would be many years later before these nebulous clouds (now known as Herbig-Haro objects) were correctly identified as radiative shocks driven by an outflow from a young star \citep{1977ApJ...212L..25S}.  The availability of new technologies and telescopes over the subsequent decades has yielded  discoveries of bipolar molecular outflows at millimeter wavelengths, high speed (hundreds of km\,s$^{\rm -1}$) optical jets in permitted and forbidden line emission, and more recently even synchrotron emitting jets associated with low mass star formation.  This extensive body of work shows that bulk flows ranging from several to hundreds of km\,s$^{\rm -1}$ are observed in these different tracers.  Although first discovered to be associated with the birth of low mass T-Tauri stars, observations over the last two decades have shown outflows to be linked with the formation of objects across the  entire mass spectrum: from brown dwarfs \citep{2005Natur.435..652W} on up to O stars \citep{2017NatPh..13..276C}.  These observations have made it clear that outflows are an essential component of star formation.

The first clues to the origin of protostellar jets  emerged from studies of outflows from low mass stars, whose radiation pressures fail by orders of magnitude to drive them at the observed thrusts (see, for example, \cite{Wu2004,2011ApJ...742...56V} on the decollimating effects of radiation in the case of massive young stars).  It was therefore natural to consider the possibility that T~Tauri jets and outflows could be magnetized winds from rotating bodies. Two kinds of magnetized rotors are possible - rapidly spinning, magnetized protostars or magnetized accretions disks out of which all stars and their planetary systems form.  While magnetic field measurements needed to confirm such a picture have been long in coming, direct observational evidence is now available using several approaches: the spectro-polarimetry of jet sources \citep{2010MNRAS.409.1347D}, synchrotron emission in outflows from some young stars %\citep{Carrasco-Gonzalez2010AProtostar, 2014ApJ...792L..18A},
and most recently the  detection of polarization of the SiO line in a jet from a low mass star arising from the Goldreich-Kylafis effect \citep{LeeCF2018}.  The observed fields range in strength from kilogauss close to the star to a fraction of a milligauss in distant parts for the outflow. Such values suggest that while magnetic forces dominate in the vicinity of the young stellar object (YSO), they are no longer dynamically important further out. This scenario is in agreement with numerical simulations \citep{2007ApJ...661..910H}.

The basic physics of how magnetized rotating stars drive winds and undergo magnetic braking as a consequence was developed by \cite{1956ApJ...124..232C} and \cite{1961MNRAS.122..473M} before being applied by \cite{1967ApJ...148..217W} who showed that the magnetized solar wind would spin down the Sun.   Magnetic field lines threading a rotating star can enforce the co-rotation of gas out to some distance along the field line - the Alfv\'en radius $r_A$ where the outflow speed equals the speed of a transverse Alfv\'en wave ($B/\sqrt{4\pi\rho}$).  In so doing, the wind is able to extract the rotor's angular momentum which is then carried out by the rotating, accelerated outflow. The lever arm of the  wind torque is, in effect, the Alfv\'en radius.  It can extend to significant distances from the body making the magnetized outflow a highly efficient mechanism for extracting angular momentum.  It is this point, more than any other, that makes magnetic fields so important in the physics of protostellar outflows and underpins their role in star formation. Moreover, the combination of magnetic fields and outflows, or more precisely winds, continue to be important in subsequent evolutionary phases as a means of removing angular momentum even when a star is on the Main Sequence \citep{2012ApJ...746...43R}. 

 Linking magnetized winds to the physics of accretion disks was first suggested in the context of accreting black holes at the centers of radio galaxies \citep{1982MNRAS.199..883B}.  It became clear that magnetized disk-wind torques, exerted by fields threading vertically through the disk and bending outwards beyond, extract some, if not most of the angular momentum from each annulus of the disk.  This, in turn, could  drive an accretion flow onto the central black hole.  When applied to circumstellar disks around young stars, magnetized disk winds (D-winds), extending over most of the disk surface, could also provide the major driver of accretion onto the central forming star \citep{1983ApJ...274..677P}. In any event this theory led rather quickly to the prediction that outflows and accretion disks must be fundamentally linked in a structure that undergoes both accretion and ejection.  In particular the ratio of the mass outflow rate to the wind driven accretion rate depends on the lever arm of the flow:  the ratio of $r_A$ to the footpoint radius $r_o$  of field line at each disk radius. Observations of a wide variety of systems typically observe a wind mass flux to accretion flux ratio, $\dot M_w / \dot M_a \simeq 0.1 $ \citep{2016ApJ...828...52W} confirming theoretical predictions \citep{PudritzNorman1986}.  Another key prediction of the theory is that jets rotate - a consequence of the fact that they carry off the disks's angular momentum. These predictions have been confirmed by a large number of 2 and 3D hydromagnetic simulations using a wide variety of initial setups (isolated cores, turbulent clumps, etc.) and numerical codes (ZEUS, FLASH AMR, RAMSES AMR, SPH MHD, etc.). A small sampling of these works starts from \cite{Shibata1985}, \cite{OuyedNature}, \cite{Krasnopolsky1999}, \cite{BanerjeePudritz2006}, \cite{hennebelle2009}, \cite{Price2012}, \cite{Stepanovs2014}, \cite{staff2015}, \cite{Tomida2015}, to the appearance of high resolution zoom-in techniques in \cite{2017ApJ...846....7K}. 
 
 While D-wind models are attractive, an alternative mechanism for generating outflows from disks was proposed in a series of papers starting with \cite{1994ApJ...429..781S} and which are comprehensively reviewed in \cite{2000prpl.conf..789S}. According to the latter model, high velocity outflows arise not from the surface of the disk but instead from the narrow annulus where the star's magnetosphere interacts directly with the inner edge of the disk, at the so-called co-rotation radius\footnote{The co-rotation radius is the region where the disk rotates, in angular terms, at the same rate as the star}. Here the magnetic field of the young star is assumed to be strong and to clear out disk matter as far as the co-rotation radius. At this radius the magnetic field switches from its inner closed configuration (associated with the star's magnetosphere) to an open configuration. This constitutes a magnetic X-point or more precisely an X ring. Matter is then launched centrifugally from the disk along the open field lines in a flow known as an X-wind. It should be emphasized that both X-winds and D-winds are driven magneto-centrifugally along open field embedded within rotating disks.  One of the key physical differences between the two models is that the X-wind only removes most of the Keplerian angular momentum from near the disk's inner edge  defined by the magnetopause (where the magnetic field of the star truncates the disk).  In contrast, the D-wind removes angular momentum from every disk radius.  The D-wind model also allows the Keplerian angular momentum at the disk inner edge to be transported to the star via gas infall along the stellar field lines.  Several mechanisms have been proposed to remove this from the star, including coupling to the disk, or by an accretion powered stellar wind. In terms of magnetic field geometries, the main differences  \citep{2007prpl.conf..261S} are where the field lines are anchored: near the co-rotation radius, the X-radius, for X-winds - or over a wide range of radii for D-winds starting from the inner disk edge. 
  
  One of the basic questions regarding the application of such magneto-hydrodynamic (MHD) wind models to protostellar disks is the degree to which magnetic fields are really coupled to the disks, given that the very low ionization of high column density disks implies that non-ideal effects are important \citep{2009ASSP...13...67K}.  A great deal of effort has been expended in developing computer codes capable of simulating these effects over the last decade \citep{2009ApJ...706L..46D, Seifried2012,  2014ApJ...796...31B, gressel2015, 2016ApJ...830L...8H}. The results of this work have now challenged the traditional idea of accretion disks as systems that are dominated by turbulent angular momentum transport, as we shall see. 

At the time bipolar outflows were discovered, protostellar disks had not yet been directly detected although their presence had been inferred from the modeling of observed spectral energy distributions using the IRAS satellite \citep{1985AJ.....90.2321R}.  Moreover, very little was known about the spins of young stars, except for the fact that a few  rotate very slowly (periods of approximately a week). The first optical images of protostellar disks, seen against the bright emission of the Orion Nebula Cluster by the Hubble Space Telescope \citep{1994ApJ...436..194O}, showed them to be at most 1$''$ in size for the nearest star formation regions. The search for disks associated with outflows, having the signatures of Keplerian rotation, required both high spatial and spectral resolution. Such capability became available with mm interferometry and an early-known jet source, HL~Tau, was shown to possess a rotating circumstellar disk, perpendicular to the jet, using CO as a tracer \citep{1987ApJ...323..294S}. Such discoveries, in fact became one of the major science drivers for ALMA, i.e.\ to discover and map protostellar disks and undertake studies of both star and planet formation within them.  The connection of outflows with disks is now being directly addressed as part of the $``$ALMA revolution$"$ wherein disks around young Class I stars, and their associated outflows (observed at millimeter wavelengths), have been resolved down to scales of a few au in nearby sources. Recent efforts have also focused on studying the formation and growth of disks in the most deeply embedded, even earlier phases of star formation (Class 0 sources), see, for example, \cite{2013ApJ...771...48T}.  

While it is generally acknowledged that magnetic fields likely play a central role in observed protostellar outflows, until recently the possibility that outflows might also control the physics of accretion disks has been largely ignored.  One reason is observational.  Magnetic fields in jets and disks are notoriously difficult to measure and the needed instruments have not been available. A second reason is theoretical. The seminal papers on accretion disk theory assumed that turbulence in disks would result in a viscous torque.  This transports angular momentum radially outwards through the disk, driving most of the disk material to accrete onto the star as the outer disk spreads out to carry off the angular momentum, see, for example, \cite{1998ApJ...501L.189A}. 
The questions here are - what is the origin of turbulence?  Is it of sufficient amplitude to support the large observed disk accretion rates?  It took nearly two decades for an explanation to emerge. The Magneto-Rotational Instability (MRI), arising from weak magnetic fields, provided a sound physical basis for turbulence in perfectly conducting, so-called $``$ideal", disks \citep{1998RvMP...70....1B}.  However, recent ALMA observations show that turbulence is not strong enough to carry away angular momentum in observed disks, at least at tens of au from a young star \citep{2015ApJ...813...99F, flaherty2017, flaherty2018}.  While the observations are consistent with numerical simulations for the mid-plane, where MRI-driven turbulence will be strongly damped, they are inconsistent with expected turbulent velocities at the disk surface, $v_{turb} \simeq 0.1-1.0\, c_s$, required to drive the observed accretion rates \citep{simon2018}. 
At the same time, numerical simulations of protostellar disks, that account for important non-ideal MHD effects, have discovered that 
the MRI is strongly suppressed in the expected dense, poorly ionized gas. It gives way instead to magnetized disk winds \citep{2014ApJ...796...31B, gressel2015}.  While further developments are needed - such as including the effects of grain evolution on non-ideal MHD process - progress in the field has been enormous.  Thus, recent advances in observations and breakthroughs in difficult numerical simulations both seem to point towards the primacy of magnetized outflows rather than disk turbulence as the driver of accretion disk physics \citep{simon2018}.  Before exploring disk winds further, it should be pointed out that alternative ways of transporting angular momentum outwards in an accretion disk have been suggested. In particular, transport can be by spiral density waves although such waves only form in disks that are a significant fraction of the mass of the central star \citep{2016ARA&A..54..271K}. Moreover embedded masses, e.g.\ newly formed massive planets, can drain angular momentum away from a disk and support accretion onto the young star; this however does not seem to be an important mechanism \citep{2011ARA&A..49..195A}. 

There is also the question of what causes the surprisingly low angular momentum of many young stars.  Early work suggested that one effect of the star's magnetosphere interacting with the disk is to spin-down the star, through magnetic braking, so that it is effectively locked to the same rotation period as the inner edge of the accretion disk \citep{koenigl1991}.
 This would explain why young stars, which show evidence of disks, rotate slower on average than their disk-less counterparts of comparable age and mass \citep{2007ApJ...671..605C}.  More recent work suggests that stellar spin may be controlled by magnetized stellar outflows that originate from the protostars themselves.  Here, the accretion of material from the inner disk flows onto the star by falling down magnetospheric field lines.  The gravitational energy released in this infall impacts the foot points of these field lines.  This ultimately gives rise to a powerful flux of Alfv\'en waves that heats the stellar corona giving rise to an accretion-powered wind - see for example \cite{Matt2012}.   As already mentioned, outflows have also been suggested to originate at the disk-magnetosphere boundary - the so-called X-wind - which are proposed to carry off the angular momentum of the inner disk before matter is accreted onto the star \citep{2000prpl.conf..789S}.  In any event, as soon as the star loses its disk, it starts to spin-up as it radius shrinks while descending its Hayashi track. Observational advances in measuring both the magnetic field and spins of young stellar objects have opened up the possibility of formulating the complete spin history of a star, the evolution of which appears to be particularly important in the early phases of star formation \citep{bouvier2014}.

Magnetically driven outflows have a large number of other important consequences for star formation.   If outflows originate from disks then the process of disk formation during gravitational collapse in a magnetized medium must be deeply connected with magnetic braking in the early collapse phase and subsequent launching of MHD outflows as the disk, or first core, starts to materialize. Disks and outflows in this view are inseparable, both being tied to collapse in turbulent, clouds.  Indeed, as has long been known, one of the central features of the observations is that outflows are most powerful during the most deeply embedded Class 0 stage when at least half the mass of the system is still in the envelope which is collapsing into the disk \citep{2009A&A...507..861J}.  Recent analytic and numerical calculations indeed show that small disks of about 20 au in size should form in magnetized collapse wherein the magnetic field is not frozen into the gas but undergoes (ambipolar) diffusion \citep{2016ApJ...830L...8H}.  This is very much in keeping with the sense of recent IRAM-PdBI observations which find that disks are typically smaller than expected from purely hydrodynamic models \citep{2019A&A...621A..76M}.  Thus, magnetic braking and the launch of outflows early in the gravitational collapse and disk formation stages of star formation, respectively, control one of the most fundamental physical properties of the initial state of a disk - its radius.  

Feedback processes arising from outflows have long been thought to play an important role in defining the mass of a star and the efficiency of star formation.  Magnetized outflows carry significant amounts of angular momentum and thrust, and are powered by the gravitational potential energy released during the collapse. In this way, outflows can act as an important, and even dominant form of protostellar feedback during star formation. This feedback can cut off the supply of infalling gas to the disk, and thereby help to limit or determine stellar mass \citep{MatznerMckee2000}.  On physical scales beyond the molecular core radius (typically $\approx$ 0.04pc), protostellar outflows could stir up the surrounding molecular cloud and drive turbulence.  This would, to some degree, stave off the formation of too much dense, star forming gas as first suggested in the pioneering paper of \cite{1980ApJ...238..158N}, and is in agreement with many current MHD simulations \citep{2016JPhCS.719a2002F}.  Feedback from protostellar outflows would then help to reduce star formation rates and efficiencies in molecular clouds addressing one of the big questions of star formation - why is the process so inefficient?   

This review examines these issues and questions from a modern perspective.  We focus on recent rapid progress in the ALMA era that has been made in the physics and observations of outflows and their role in star formation.  ALMA has provided a radically new capability to study disks and their outflows.  Most of our attention is on observational and theoretical results that have arisen since 2014 post the publication of Protostars and Planets VI. We first review the observational advances in measuring magnetic fields in stars, jets, and disks, and the properties of outflows (shocks, rotation, etc.).  We then go on to examine the theoretical and computational results on how outflows are launched, and their significance in various aspects of star formation.  These latter angular momentum related topics include outflows and their effects on - gravitational collapse, disk formation, disk evolution, and protostellar spin.  We then discuss the feedback properties of outflows including their effects on stellar mass and molecular cloud turbulence.  We address these issues in the context of both low and high mass star formation.  The latter subject brings up the difficult question of how powerful protostellar radiation combines with  magnetic outflows to determine conditions in massive star forming regions.  Finally, since stars and planets form in disks, both must be affected by outflow physics and so we address how planet formation could be affected by outflows.   The reader may consult reviews of earlier material in \cite{2007prpl.conf..231R}, \cite{2007prpl.conf..277P}, \cite{2014prpl.conf..451F}, \cite{2014prpl.conf..173L} and \cite{2016ARA&A..54..491B}.
  
\section{OBSERVATIONAL OVERVIEW}
It is remarkable that although magnetic fields are thought to collimate YSO jets, very few observations of field strength or direction are known. In contrast, through optical and near-infrared emission line imaging and spectroscopic studies, many jet parameters can be derived, e.g. neutral density, ionization fraction, temperature, jet opening angle, radial velocity, etc. With the addition of multi-epoch imaging, quantities such as the tangential velocity of knots, post shock cooling times, etc., can also be found \citep{2014prpl.conf..451F}, allowing a full 3-D kinematic study of outflows.  We begin our overview from the largest outflow scales, and gradually focus down to accretion and outflow from the protostellar surface. 

\subsection{Do the Magnetic Fields of the Parent Cloud or Cores Determine the Outflow Direction?}

It is now known that outflows from young stars extend to parsec and even tens of parsec scales, i.e.\ lengths comparable to the size of their parent molecular cloud \citep{2016ARA&A..54..491B}. This is not surprising when one considers that it takes approximately 1 million years \citep{2009ApJS..181..321E} for a solar mass young star to go through its outflow phase (i.e.\ from Class 0 to Class 2) and that typical outflows have velocities of several hundreds of kms$^{\rm -1}$. Therefore an obvious starting point is to consider whether outflows are somehow aligned, either directly or indirectly, with the magnetic field of their parent cloud. Conceivably this could be
through some sort of guiding action by the cloud's magnetic field or alternatively, since accretion disks and outflows are perpendicular, by the field determining the disk's orientation. We can immediately dismiss the first suggestion since outflows are not only highly supersonic with respect to their internal and external sound speeds but also jet velocities are much higher than their internal and external Alfv{\'e}n speeds. The second idea however is worth exploring. Despite low ionisation levels, neutral and ionized matter in a molecular cloud are strongly coupled through collisions, and thus matter is preferentially accreted along magnetic field lines. In turn one might naively expect disks to form in the perpendicular plane.  

As described elsewhere in this volume, magnetic field directions in a cloud can be obtained either through polarized absorption of continuum light by dust grains, when the optical depth is low, or through polarized emission at infrared/mm wavelengths from dust grains, when depths are high. Pioneering studies by \cite{1987IAUS..115..255S} suggested there might be a relationship in the sense that the directions of outflows could follow that of the magnetic field in the surrounding cloud. Number statistics at the time however were very poor and subsequent studies have suggested there is no such correlation \citep{2004A&A...425..973M, 2011ApJ...743...54T} or at best present only for the youngest protostars. 
Of course expecting such a correlation assumes disks form only by the collapse of material along the field lines and subsequent distortion of the field into the classical `hourglass' pattern as gravity draws material inwards. Many studies however have shown that magnetic braking in such a case is far too effective and that disks, at least of the large sizes observed, should not form even when non-ideal damping MHD effects are taken into account \citep{2009ApJ...706L..46D, 2012ApJ...757...77K}. Moreover we also have to be careful in looking for a correlation with the large scale ($\approx$\,1\,pc) magnetic field direction in the parent cloud and the outflow when perhaps we should be searching for a link with the magnetic field direction in the smaller parent core and envelope. As polarization in the latter can only been measured when the optical depth is sufficiently high, and the scales we wish to investigate are at most a few thousand au, one has to turn to millimeter interferometry of younger embedded protostars. Here initial results \citep{2014ApJS..213...13H} do not suggest any correlation and, in fact, if anything a tendency for outflows and magnetic fields in the cores {\em to be orthogonal}. We note however that a smaller study that used a more carefully selected sample of disks, came to the opposite conclusion, namely that there is a good correlation between outflows and magnetic field geometry on the scale of cores in class 0 sources \citep{chapman2013}.  Overall, a lack of correlation is understandable in the context of more up-to-date MHD simulations which show that disks, similar in diameter to what is observed, are only formed if there is {\em misalignment} between the rotation axis of the core and the ambient magnetic field {\em providing the field is weak} \citep{2012A&A...543A.128J, 2013ApJ...767L..11K}. Moreover the effectiveness of magnetic braking is also in line with the recent study of \cite{2018A&A...616A.139G} who has shown there is increased likelihood of alignment between the outflow direction and the orientation of the ambient magnetic field if there is no small-scale multiplicity and no large disks on $>$\,100\,au scales. 
In summary then we do not expect, or find, any strong correlation between outflow orientation and the direction of the ambient magnetic field on scales of several hundred au to parsecs. 

\subsection{What Do Observations Tell Us About Magnetic Fields in Jets Far from the Young Star?}

It is well known that most of the radiation emitted by jets from young stars comes from the cooling zone behind shocks. The properties of such shocks, in the absence of a magnetic field, have been modelled for a number of years and depending on the shock velocity various emission lines and relationships between the fluxes of such lines, are expected in the post shock zone. At optical wavelengths for example, we expect to see both lines from neutral, e.g.\ from H, O, and singly ionized species, e.g.\ S$^{\rm +}$.  As jets from young stars, or more precisely their atomic/ionized component, typically travel at hundreds of kms$^{\rm -1}$, we would also predict the presence of highly ionized species, for example O$^{\rm ++}$, in the post shock cooling zone if most of the outflow energy is converted into radiation at a single shock. In reality, except for the tip of the largest bow shock shaped features (e.g.\ in HH\,34) such emission is not observed. This is because the shock fronts seen in jets are so-called $``$working surfaces$"$ in which the shock velocity is determined by {\em the difference in velocity} between consecutive flows from the young star \citep{2002ApJ...565L..29R}. The line emission in this case is then consistent with a shock velocity of several tens of kms$^{\rm -1}$ rather than a few hundred kms$^{\rm -1}$, i.e.\ in line with the velocity differences. Recent 3-D simulations, with realistic cooling, show the rich structure that such variations can lead to, see, for example, \cite{2017ApJ...837..143H}. What is important to emphasise is that the results of such simulations are in very good agreement with what is found through observations, for example, with HST.

Of course in a standard, i.e.\ adiabatic, strong shock the density of the flow increases by a factor of 4 and the velocity correspondingly decreases by the same factor. Here we are referring to velocities in the shock frame and by the term strong shock, we mean one in which the incoming velocity V$_{\rm s} \gg {\rm c}_{\rm s}$ where c$_{\rm s}$ is the sound speed. In other words M$_{\rm s} \gg $\,1 where M${\rm_{s} = V_{s}/c_{s}}$ is the Mach number of the shock. In a radiative shock, due to the loss in energy, gas in the post shock zone is compressed over and above that expected for an adiabatic shock. The increased density leads to further collisional excitation and strong line emission is produced from oxygen, hydrogen, sulphur, etc. Gas density is then increased further in order to maintain pressure and effectively these lines act as thermostats keeping the temperature around 10$^{\rm 4}$\,K.

Unfortunately the presence of a magnetic field in a jet cannot be determined in a straightforward manner through, for example, observable Zeeman splitting of its emission lines. If a magnetic field is present however, in particular one parallel to the shock front, or with a significant component that is parallel, then clearly the field will resist compression in the cooling zone and, in turn, this will have an impact on the shock’s emission properties, e.g., observed line ratios. However, as pointed out many years ago by \cite{1994ApJ...436..125H}, interpreting the results is complicated. This is because observable line ratios are degenerate with respect to the field's strength as spectra from low velocity shocks, without a magnetic field, resemble those from higher density, higher velocity shocks with a field. Breaking this degeneracy requires additional information which may be present. For example if the shock is well resolved spatially and is bow shaped, then the extent of the [O\,III] emission near its tip, can be used directly to infer the shock velocity at the apex. In addition the H$\alpha$ flux from the bow, assuming it is clearly separated from the cooling zone, and the [SII]$\lambda\,6716/\lambda\,6731$ line ratio can be used to determine both the pre-shock density and the post-shock compression respectively. This in turn reveals the strength of the pre- and post-shock fields in the shock plane. In this way \cite{1992ApJ...399..231M, 1993ApJ...410..764M} found weak fields, of around 30\,$\mu$G, in the gas ahead of bow shocks in the HH\,34 and HH\,111 outflows. Compression of the field then resulted in around 1\,mG in the cooling zone. It should be said that this equates to an Alfv{\'e}n speed of around 10 kms$^{\rm -1}$ in most of the observed post-shock zone, i.e.\ comparable to the sound speed but orders of magnitude smaller than the velocity of the jet itself. Put another way, in these outer regions the magnetic field is no longer dynamically important although the situation may be very different close to the source. 

A variation on the above method to measure the magnetic field using shock physics has been proposed by \cite{2015ApJ...811...12H}. Here the degeneracy is broken by measuring the extent of the cooling zone which is  
obtained from the distance, allowing for projection effects, between the Balmer emission lines and the forbidden lines. Applying the method to existing data for a bright knot in the HH\,111 jet, they obtain a relatively low Alfv{\'e}nic Mach number, ${\rm M_{A} = V_{S}/V_{A}}$, indicative of super-magnetosonic velocity perturbations in the jet.

Another important effect that magnetic fields in jets have is on softening the effects of a shock. In a standard non-magnetic shock, the transition from pre- to post-shock gas conditions is sudden and the width of the shock is a few mean free paths. The presence of a magnetic field effectively allows energy to be transmitted upstream ahead of the shock, accelerating it, and giving rise to a gentler profile in which the heating of molecules and atoms is more gradual \citep{1997A&A...326..801S}. Such shocks are known as C (Continuous) shocks as opposed to their non-magnetic J (Jump) shock counterparts. One effect of the more gradual changes in properties in a C-shock is that molecules, e.g. H$_{\rm 2}$, may not be dissociated even at shock velocities where this might be expected \citep{2000A&A...359.1147E}

The presence of magnetic fields can also be indirectly inferred from jet rotation. The theory is that the wind leaving the accretion disk is collimated into a jet and focused on scales comparable to the Alfvén radius. Up to such distances the ionized wind/jet material is forced to corotate with an angular velocity equal to that of its foot-point. Since the wind is expanding away from the star, along magnetic field lines, it gathers angular momentum as it does so. Through the magnetic field there is then a back reaction on the disk, forcing material in the disk to lose angular momentum and continue to spiral inwards. It is therefore possible for the wind/jet to carry away say 10\% of the mass that flows through the disk but practically all of the angular momentum of the remainder. If this scenario is correct, then jets must rotate. The amount of rotation depends on where the jet is launched from. For example, if the X-wind model is correct and the jet comes from close to the disk’s co-rotation radius (typically at a few stellar radii), then the expected spin of the jet should be small. If, on the other hand, most of the material comes from further out, say at ($\sim$ 1\,au), then we expect any measured jet rotation to be corresponding higher. Here we are assuming that the magnetic level arm is similar in both cases, i.e. the ratio of mass outflow to mass accretion is the same. 

     The search for jet rotation initially used the Space Telescope Imaging Spectrograph (STIS) on the HST \citep{2002ApJ...576..222B, 2007ApJ...663..350C} either in the optical or UV. Here the high spatial resolution of this instrument, of order 0.1 arcseconds was particularly important. The reason for this is that if jets are launched within a few au of a young star, then one might expect a good place to look for signatures of rotation is close to the source, say within a few hundred au. In this region the jet has not had time to interact with its surroundings, an effect that might mask, or even mimic, rotation. In this region however, the jet width is at most a fraction of an arcsecond for the nearest young stars \citep{2000A&A...357L..61D}. Thus, high spatial resolution, such as afforded by HST, is needed. Moreover, if a typical jet is launched with a velocity of a few hundred km s$^{\rm -1}$ then at most we expect differences in radial velocity of say a few tens of kms$^{\rm -1}$ from one side of the jet with respect to the other in a slice transverse to the outflow axis. As shown for example by \cite{2000ApJ...537L..49B}, a transverse cut through a jet does not show a $``$top-hat$''$ velocity profile, even in the absence of rotation. Instead such position velocity (PV) diagrams show a smoother decrease towards the jet edges, i.e.\ the poloidal velocity gradually decreases away from the jet axis. The search for a rotation signature is then a hunt for a lopsidedness, i.e.\ a lack of axisymmetry, in this profile. 

These observations are extremely challenging however, even with STIS, given the angular widths of the nearest jets and the spectral resolution required.  Nevertheless, observations of jet rotation in theory could help discriminate between different MHD jet wind launching models. One other obvious constraint that we might also reasonably impose is that any evidence for rotation must be consistent with the sense of rotation of the underlying accretion disk and that the opposing (red and blue-shifted) jets in a bipolar outflow must have opposing helicity. With these constraints in mind, results from optical/UV observations using HST have been something of a mixed bag. Moreover, even when rotation has been claimed, alternative explanations such as asymmetrical shocks \citep{2016ApJ...832..152D} and precession \citep{2006AIPC..875..285C} have been proposed to account for the observed effects. 

An alternative observational approach in the search for rotation has however emerged in recent years. Instead of looking for a rotation signature in the atomic/partially ionized jet, studies are now being undertaken of the molecular outflow in the millimetre band using high spatial and spectral resolution interferometers such as ALMA. Although the available spatial resolution is not as good as HST (although resolutions of 0.02" have been achieved for ALMA observations of HH212, \cite{2017NatAs...1E.152L}), molecular jets tend to be wider than their atomic counterparts close to their source and the spectral resolution of an instrument like ALMA is much higher than STIS. Thus, it is possible to search amongst molecular jets for rotation signatures. 

\begin{figure}[h!]
\begin{center}
\includegraphics[width=18cm]{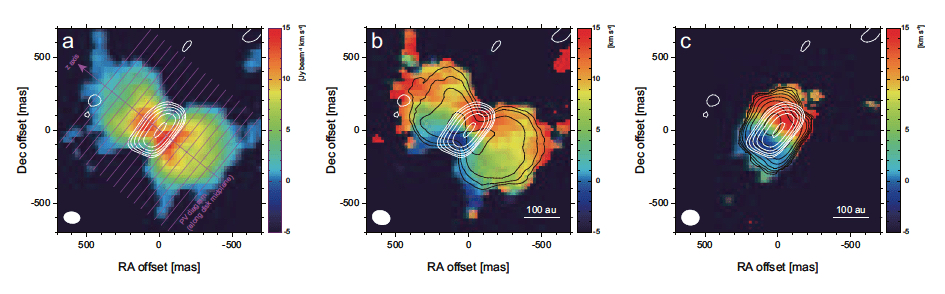}
% This is a *.eps file
\end{center}
\caption{ALMA observations of rotation in the bipolar outflow driven by Orion Source I, a high mass YSO candidate. White contours are 490 GHz continuum maps. (a) moment 0 (colour; integrated intensity) map of 484 GHz emission in Si$^{18}$O; (b) moment 0 (black contours) and moment 1 (peak velocity; color) map of the 484 GHz Si18O line; (c) moment 0 (black) and moment 1 (color) maps of the 463 GHz H2O line. These latter two plots show rotational velocity gradients of the outflow- water line emission being more compact.  From \cite{2017NatAs...1E.146H} reproduced with permission \textcircled{c} Springer Nature. }\label{OriSourceI}
\end{figure}

Figure ~\ref{OriSourceI} shows an outstanding example of a rotating, large scale molecular outflow observed by ALMA \cite{2017NatAs...1E.146H}- a source whose rotating outflow was first discovered by SiO maser observations \cite{Matthews2010}. Observed at resolutions of about 50\,au, observations were made in Si$^{18}$O (484 GHz) and H$_2$ O (463 GHz) lines of an outflow from a region of massive star formation in the famous Orion nebula (KL region).  The former traces emission from the outflow while the latter traces emission from a dense rotating disk.  The outflow speed is $\sim 18 $ km s$^{-1}$ on an outflow scale of 1000\,au.  We are afforded an almost edge-on view of the system.  Using observations of the rotation curve of the gas in the disk, \cite{ginsburg2018} were able to estimate the mass of the central star (SrC 1) to be $ \sim (15) M_{\odot} $. Both these maps (Figures \ref{OriSourceI}b and \ref{OriSourceI}c) show velocity gradients along the disk plane.  

The authors note that other possible explanations of this velocity gradient such as disk warping and other effects can be ruled out for this source making it likely to be an example of a rotating molecular outflow.   The resolution is not sufficient to distinguish Keplerian vs constant angular momentum ($v_{\phi} \propto r^{-1}$ ) for the disk.  However the specific angular momentum of the outflow is approximately constant.  With a measured value of this wind angular momentum of 400-600  km\,s$^{-1}$\,au, this corresponds to a source radius on the disk (where centrifugal and gravitational forces balance) of 21-47 au.  The observations suggest that this is almost certainly evidence for a disk wind being accelerated off the disk far from the star or star-disk magnetosphere, and that the wind material is not swept up through entrainment by a narrow jet from the surrounding gas envelope.  
There are however a number of caveats worth pointing out about Source I and its immediate vicinity. First, like most massive stars, Source I is in a cluster that may not only be dynamically active, but which can cause confusion when trying to untangle the effects of one outflow from another. In particular, while previous work suggests a very young jet-like outflow from Source I in SiO and CO, this interpretation is less clear in recent ALMA data \citep{2017ApJ...837...60B}. 
Another point that should be stressed is that while our understanding of how high mass stars form seems, in the main,
to resemble our picture for low mass star formation other physical processes must come into play. For example, as shown by simulations, increasing uv radiation and its ionizing effects, must be taken into account \citep{2015ASSL..412...43K}.

\begin{figure}[h!]
\begin{center}
\includegraphics[width=15cm]{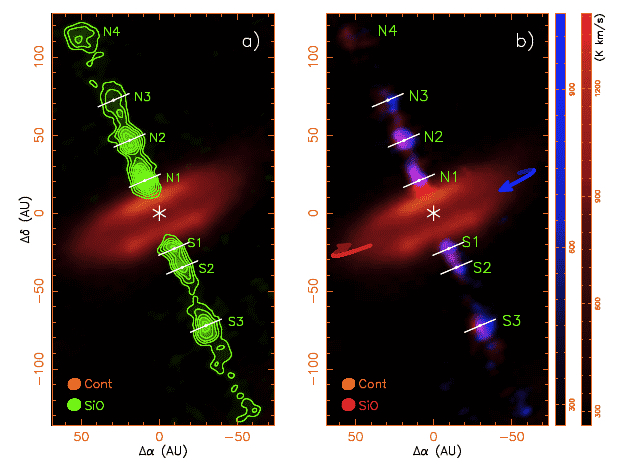}% This is a *.eps file
\end{center}
\caption{ALMA observations of the rotating outflows of the class 0 protostar HH\,212. Shown is a zoom-in to the innermost part of the jet in SiO within $\approx$ 120 au of the
central source, at a  resolution of $\approx$ 8\,au on top of the continuum map of the
disk. The maps show the intensity (in unit of K\,km s$^{\rm -1}$) integrated over certain velocity range. (a) A chain of new
knots trace the primary jet emanating from the disk, observed in SiO. (b) Blueshifted and redshifted SiO emission of the jet plotted with the
continuum emission. Direction of rotation is shown with arrows. From \cite{2017NatAs...1E.152L} reproduced with permission \textcircled{c} Springer Nature.   } \label{HH212}
\end{figure}

Figure \ref{HH212} shows ALMA observations of an SiO jet in the famous class 0 outflow source HH\,212, located in the  Orion L1630 cloud at a distance of 400 pc.  The importance of this source is that it allows us to probe the very earliest stages of the launch of outflows and their connection with disks.  ALMA observations  \cite{2017NatAs...1E.152L} reveal knots in the highly collimated jet within 120\,au of the rotating disk (shown in red), down to 10\,au scales. Of central importance is the 7$\sigma$ detection of velocity gradients across these knots, perpendicular to the outflow.  The sense of the gradient is the same for all of the knots, ruling out random fluctuations in the jet.  The mean specific angular momentum in the jet is $l_j \sim 10.2 \pm 1.0$ au\,km\,s$^{\rm -1}$.  By assuming the conservation of angular momentum along streamlines, and projecting this back to a source radius on the disk \citep{2003ApJ...590L.107A}, one finds that the footpoint of this flow is $r_o \approx 0.05$\,au, originating from the innermost regions of the disk.   This is consistent with an origin from the innermost zones of the disk, from either a close-in disk wind, an X-wind or even a stellar magnetospheric wind.  

A number of other rotating outflows have now been found, including sources such as DG Tau \citep{2011A&A...532A..59A}, Orion BN/KL Source I \citep{2016ApJ...824...72C}, TMC\,1A6 \citep{2016Natur.540..406B} with flows originating from extended disk scales, and the rotating and possibly precessing molecular outflow in HH30 \citep{louvet2018}.  In the case of the DG Tau source observed with the Plateau de Bure interferometer, the most likely explanation for the highly collimated jet seen in Fe lines is a quasi-steady centrifugal MHD disk wind ejected over 0.25-1.5\,au and/or episodic magnetic tower cavities launched from the disk. The rotation seen in SMA observations of the outflow in Orion BN/KL suggests that the jet launching footpoint on the disk has a radius of 7.2-7.7\,au.  ALMA observations of the CO outflow in the TMC1A6 outflow indicate a source radius of 25\,au from the disk, ruling out X-wind or stellar wind sources for this molecular outflow.  In a similar vein, \cite{Zhang2018} have observed the NGC 1333 IRAS 4C outflow in the Perseus Molecular Cloud with ALMA, tracing CCH and CS emission in an outflow cavity. Outflow rotation is detected from 120 - 1400\,au above the disk, and its velocity is highly symmetric about the outflow axis.  A flat distribution of specific angular momentum profile is observed in the outflow, with a mean value of 100\,au km\,s$^{-1}$. Projecting this back onto the disk gives a launch radius of 5-15\,au, indicating a wind originating from large portions of the disk. The wind may reach out to the centrifugal radius of the disk ($r_c \simeq$ 30\,au), so further high resolution observations are needed.  

These various examples suffice to make the point that ALMA resolves rotating molecular outflows that originate from large sections of their disks, and perhaps out to their edges. 

%Point made by John Bally regarding "onion-like" velocity profile in this section is already made in the earlier piece regarding results from STIS.   

\subsection{Outflows and Disk Formation}

The observational evidence shows quite clearly that stars are born within surrounding disks or 
disk-like structures that, of course, are the birthplaces of
planets. Nevertheless,  there is still considerable debate in the theoretical models as to how and when
protostellar disks first appear and what their basic physical properties are in these early phases
(see the article by Würster and Li, this volume). Certainly multiplicity, across the stellar mass spectrum, is likely to 
be important in how disks evolve, and to effect outflow production. For example most class-0 low mass stars are in multiple systems and the level of multiplicity appears to decline as such systems evolve \citep{2012Natur.492..221R,2015AJ....149..145R}. This immediately suggests that dynamical, so-called N-body interaction, must be taken into account as it can lead to disk truncation and warping, precession of outflows, etc. Despite such complications, it is important to examine what effects magnetic fields might have on disk and outflow evolution. How influential magnetic fields are, can be gleamed 
from examining their structure, i.e. geometry, during the early stages of star formation. 

The study of magnetic field structure in dense molecular clouds relies on the fact that elongated and spinning dust grains are aligned perpendicular to magnetic field lines.  While radiation effects produce the spin needed for alignments by magnetic field, it is important to note that as asymmetric dust grains scatter or absorb photons they also undergo radiative torques which can be stochastic or regular  \citep{2009ApJ...697.1316H}.  These radiative torques leading to grain alignment would be minimal in regions of sufficiently high optical depths. Polarization observations of background star light passing through cores in molecular clouds are not possible at optical or infrared wavelengths because of the high column densities of such regions.  However, thermal emission from these spinning grains can be detected at millimeter and submillimeter wavelengths and is weakly polarized in a direction perpendicular to the field lines  \citep{2009ApJS..182..143M}. Using these techniques, the relative orientation of bipolar outflows with magnetic field directions towards 16 class 0 and I outflows were measured  \citep{2013ApJ...768..159H}. These high resolution submm observations were carried out at the CARMA. The results showed that magnetic fields on ~ 1000 AU scales are consistent with models in which outflows and magnetic fields are either randomly aligned, or preferentially perpendicular.  These results may also be interpreted as the random alignment of disks with magnetic fields - as would arise if turbulence plays an important role in disk formation (see review by Hull and Zhang in this volume).  

\begin{figure}[h!]
\begin{center}
\includegraphics[width=18cm]{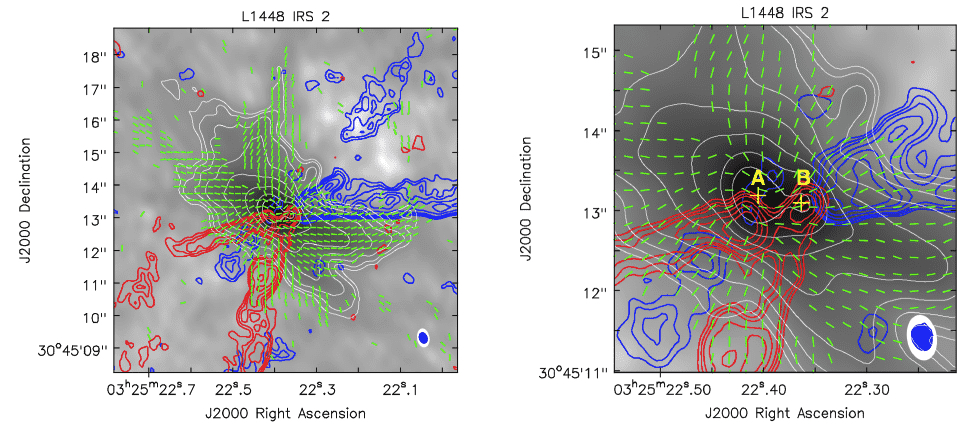}% This is a *.eps file
\end{center}
\caption{ Magnetic field morphology around L1448 IRS 2. Left: The green vectors indicate the inferred magnetic field direction. The gray scales and white
contours present the total intensity distribution with levels of 2 to 129 times 0.1 mJy
beam. The blue and red contours are CO 2-1 intensity distributions integrated in velocity ranges of -8:0
to 2.0 and 7.5 to 16.0 km s$^{-1}$, respectively. Right: Zoomed-in central region. The synthesized beams of the CO and the continuum data are marked
in the bottom right corner in blue and white, respectively. From \cite{kwon2019} reproduced with permission \textcircled{c} AAS.} \label{L1448IRS2}
\end{figure}

Probing the actual geometry of outflows and magnetic fields on disk scales is now possible with polarimetric observations at mm as well as mid-infrared wavelengths \citep{2019ApJ...872..187C, 2019ApJ...870L...9J}.  Observations of L1448 IRS 2 in ALMA band 6 were carried out by \cite{kwon2019}.  Dust polarization can be affected by radiation anisotropy as well as self-scattering, but these it is argued may be minimal in this source.  Figure \ref{L1448IRS2} shows that the magnetic field takes on the classic hour glass form expected for contracting magnetized disks threaded with an ordered field.  The field is aligned with the outflow in the vertical direction, but switches to perpendicular alignment at the disk.   The latter effect indicates the presence of a strong toroidal field at the disk.   If this magnetic field interpretation is correct, then the field has been strongly wound up in the disk. The Davis-Chandrasekhar-Fermi (DCF) method was used to measure the field strength in the disk region of 720 $\mu$G, although some caution needs to be exercised in applying this to regions in which turbulence may not be strong.  This strong field would brake the disk quickly; in 1400 years. The ordered outflow and presence of a large disk perpendicular to it might suggest that magnetic braking effects have been minimized - perhaps by ambipolar diffusion. \cite{2019ApJ...872..187C} present the best magnetic field 
map of the OMC1 core and outflow region to date. Again a classic hourglass pattern is seen.   

Perhaps the largest sets of high resolution images that we now have of protostellar disks is a consequence of the Disk Substructures at High Angular Resolution Project (DSHARP) campaign at ALMA.  The initial survey of 20 protostellar disks at a resolution of 5au FWHM provides the first look at the small-scale features in disks that are directly relevant to the planet formation process. It must be noted that this is a diverse collection of disks that are typically much larger than the bulk of disks now turning up in other surveys. They are images of Class II systems (ie disks well beyond the formation phase), and can be regarded as protolanetary disks (disks in earlier phases of star formation - the so called Class 0 and I phases, are denoted as protostellar disks).   These images show a variety of rings and gaps, amplitudes, and disk sizes for targets with a range of stellar and disk properties \citep{2018ApJ...869L..41A}.  The ubiquitous ring and gap structures have no obvious spacings that are connected with properties of the host star.  The ring systems may arise through modulations of disk pressure that trap dust grains. Specifically, the observed system rings have been modeled as arising by dust trapping in axisymmetric pressure bumps \citep{2018ApJ...869L..46D}.  If the turbulence in disks is relatively weak, then dust can be strongly confined in pressure bumps, while strong turbulence compromises the trap.  Theoretical modeling of the ALMA ring data suggests a lower limit to the turbulent amplitude expressed in the standard turbulent viscosity parameter  ($\alpha$) which takes a lower bound of $\alpha \ge 5 \times 10^{-4}$. Without this level of turbulence, the rings would be even sharper features than they already are.  It is possible that these pressure bumps themselves are produced by planets.  In any case,  the model constraints on the turbulence level is of interest in comparing the relative role of turbulent versus disk wind torques in disks.  

\subsection{What Radio Emission from Outflows Can Tell Us}

Observations of YSO jets at radio wavelengths (around a few centimetres) have shown them to primarily emit thermal radiation (i.e.\ radiation with a positive spectral index $\alpha$ where the flux at frequency $\nu$, $S_\nu \propto \nu^{\alpha}$). Moreover, typical flux densities are very low and usually less than 1 mJy \citep{2018A&ARv..26....3A}. Such emission is un-polarized and is often elongated in the direction of the known atomic/molecular outflow. Until recently the amount of information that could be obtained from radio maps was limited, primarily due to the lack of sensitivity of radio interferometers, nevertheless they provided upper limits to the diameter of jets and have proven that they must be collimated on scales less than 50\,au in agreement with high spatial resolution HST data \citep{2004AJ....127.1736R}. 
In the past few years however radio astronomy has undergone a revolution that is having a major impact on our ability to image outflows from young stars. This revolution is akin to the replacement of photographic plates in optical astronomy (with quantum efficiencies (QE) of a few per cent) by the CCD (QE $\approx$ 70\% or more). 
In the past traditional radio interferometers, such as the Very Large Array (VLA) in New Mexico and the high spatial resolution Multi-Element Radio Linked Interferometer (MERLIN) centred on Jodrell Bank, used microwave links between the various telescopes to transmit wavelength, amplitude and phase information. These links were of limited bandwidth
and correspondingly only a narrow radio continuum could be observed at any one time. Such old links however have now been replaced by fibre, allowing considerably larger observing bandwidths that are used in conjunction with much-improved correlators, lower noise amplifiers, and increased computer processing power. The net result is an enormous increase in radio continuum sensitivity (typically by an order of magnitude or more). For example the VLA in standard continuum observing mode had 4 sub-bands of 50 MHz each but the JVLA has 2 sub-bands 2 GHz wide!  Of course this has a dramatic impact on radio studies of jets from young stars.  Complete radio surveys of YSOs in nearby star formation regions, for example, have become possible for the first time \citep{2015ApJ...801...91D,2018arXiv180602434T}. 

A second strand to the radio astronomy 
$``$revolution$"$ is the opening up of the low frequency radio spectrum with the advent of instruments such as the Low Frequency Array LOFAR \citep{2013A&A...556A...2V} radio telescope. LOFAR is allowing us to explore jets from young stars in a new band (at meter wavelengths) and, as described below, could reveal any non-thermal emission.
	A particular advantage of the radio band is that it can potentially provide very valuable information on magnetic fields in YSO jets.  Theoretically we only expect magnetic fields to dominate the dynamics of an outflow close to the source, i.e. within 100\,au, and in high velocity post-shock regions where the field is compressed and amplified \citep{2007ApJ...660..426H}. Thus we have a higher chance of measuring jet fields by studying the zone surrounding the young star at high angular resolution and the brightest knots (compressed zones) on more extended scales. But how might we do this? One powerful method is hinted at by the fact that a small number of jets/outflows have non-thermal radio spectra as demonstrated in the case of more massive young stars such as Herbig-Haro 80/81 \citep{2010Sci...330.1209C} but also in their lower mass counterparts \citep{2014ApJ...792L..18A}. Such radiation 
	appears to come from high-energy electrons (i.e.\ it is gyro-synchrotron or synchrotron emission) and, as discussed below, it may be common at very weak flux levels in “standard” low luminosity outflows but nevertheless detectable with the new suite of radio interferometers such as e-MERLIN, the Jansky VLA (JVLA) and at low frequencies (e.g.\ using LOFAR).

\begin{figure}[h!]
\begin{center}
\includegraphics[width=12cm]{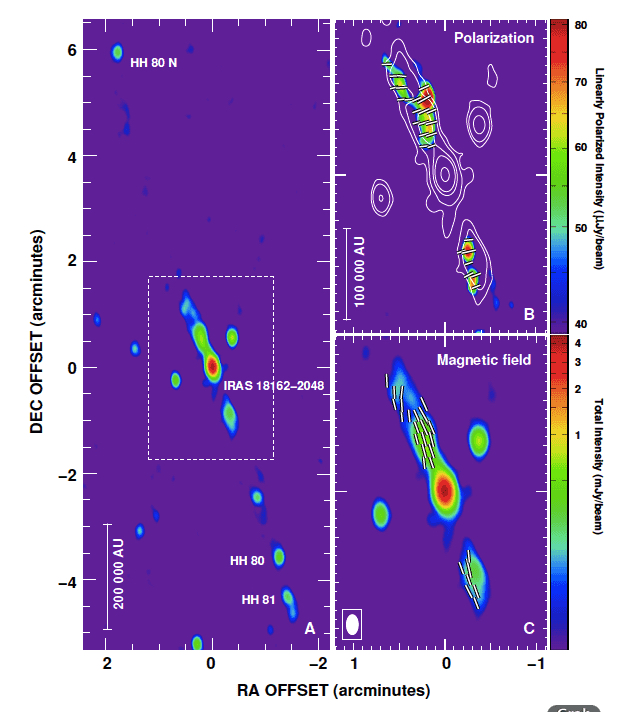}% This is a *.eps file
\end{center}
\caption{High resolution, 6 cm radio continuum VLA images of the jet HH80-81.  The knots, 0.5 pc from the source are linearly polaized indicating non-thermal synchotron emission from the jet. A. radio continuum image of the whole jet; B. Linearly polarized continuum intensity image; C. Apparent magnetic field directions superimposed on continuum map.  From \cite{2010Sci...330.1209C} reproduced with permission \textcircled{c} AAAS. } \label{HH80-81}
\end{figure}

Figure \ref{HH80-81} shows the radio continuum map of HH\,80-81 \citep{2010Sci...330.1209C}. The analysis of the synchotron emission yields a field strength of  $\sim 0.2 $ mG in the jet.  The field is parallel to the jet axis, and the increase of polarized emission toward the limb of the jet is evidence for a collimating toroidal field.  These characteristics are very similar to those of AGN jets, providing good evidence to support the idea that the jet mechanism is universal.   

As stated previously, we primarily expect to detect free-free emission in the cm (1-10 GHz) band. However, non-thermal emission is observed on occasions and this opens up the intriguing possibility of directly measuring both the magnetic field’s strength (using its intensity) and direction (using polarization). Such measurements, although difficult, could provide the $``$missing link$"$ needed to properly evaluate the importance of magnetic fields in jets, as essentially all other parameters are known. The improved sensitivity and high resolution of e-MERLIN and the JVLA may allow us, in a systematic way, to study this type of emission as faint non-thermal radiation (arising from electrons being accelerated to relativistic speeds near shocks) mixed in with free-free emission \citep{2016ApJ...818...27R}. 

\subsection{Magnetic Fields Close to Young Stars}

Magnetic fields are thought to play a very important role in accretion onto young stars and the evolution of  their outflows. In particular circumstellar magnetic fields are considered to be sufficiently strong to clear the central region of  a young star's accretion disk and force the star to co-rotate with the disk’s inner edge \citep{2013EAS....62..169F}. Moreover, at the same time as they funnel material onto the star, they may also help remove excess angular momentum thereby allowing accretion to proceed and the star to contract to the main sequence. Magnetic fields may even play a role in planet formation since the inner gap they generate could allow hot Jupiters, which have already been discovered around T Tauri stars \citep{2016Natur.534..662D}, to survive without them being accreted by their parent YSO \citep{2009ApJ...702L.182A}. 

\begin{figure}[ht]
\begin{center}
\includegraphics[width=18cm]{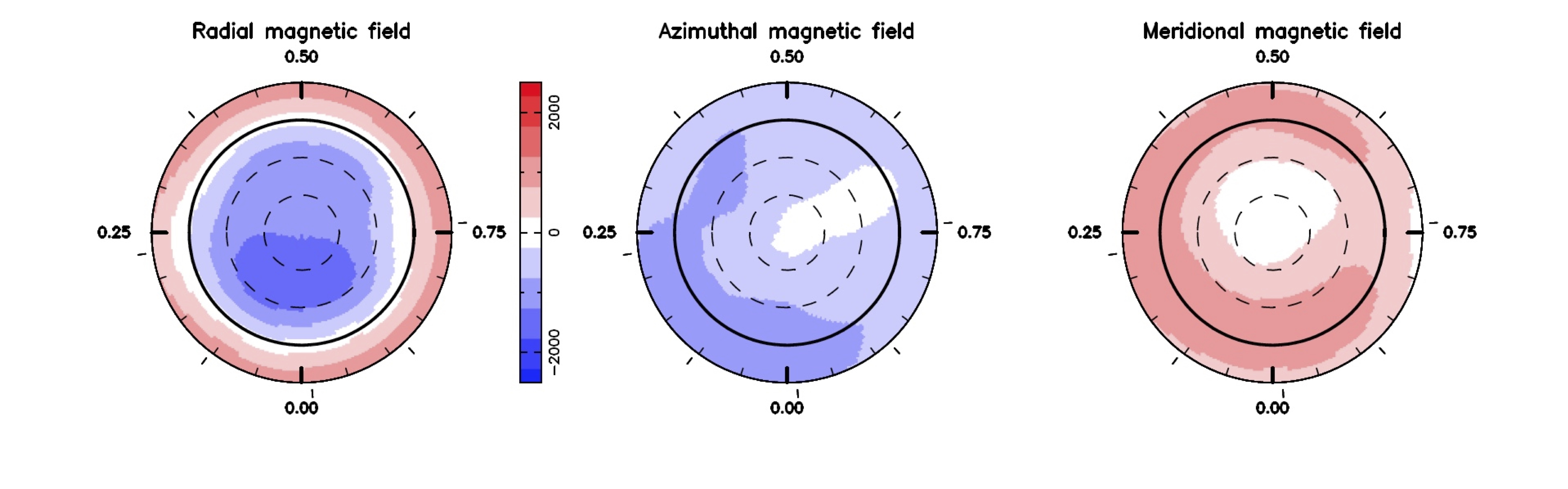}
\end{center}
\caption{Magnetic field map of the surface of the known jet source AA Tau.  The field strength, shown by the colored scale, is in gauss for the radial, azimuthal and meridional components which are shown as a function of stellar phase. From \cite{2010MNRAS.409.1347D} reproduced with permission \textcircled{c} OUP.  }\label{AATauMag}
\end{figure}

It is only in recent years that we have been able to determine the geometry and strength of the fields in classical T Tauri stars \citep{2007ApJ...664..975J,2012MNRAS.425.2948D}. Such studies have revealed some surprises, including the fact that the weakest fields have the most complex geometries, i.e. they are not simple dipoles, and that these tend to be associated with the development of a substantial radiative core (Figure \ref{AATauMag} Left). Moreover these fields are found to vary, which with the addition of their complex geometry, suggests they are generated through dynamo action and are not fossil fields as has been assumed in the past. 
Determining the magnetic field direction and strength can be done using Zeeman Doppler Imaging or ZDI \citep{2011IAUS..273..181M,2012ApJ...755...97G}, i.e. by spectro-polarimetry using instruments like HARPS-Pol at ESO. This involves measuring the circular polarization signal (Stokes V parameter) in both photospheric absorption lines and accretion dominated emission lines over several rotation cycles. The signal in accretion lines is usually so strong that measurements can be made using individual lines (e.g. He I 5876) although it should be stressed that this represents the field in the vicinity of an accretion $``$hotspot$"$. In contrast the signal in magnetically sensitive photospheric lines is so weak that cross-correlation techniques over many lines have to be employed to get an average Zeeman signature \citep{2012ApJ...755...97G}. Combining both sources of data together, and by monitoring how the Stokes V parameter changes as a function of rotational phase, it is then possible to map the magnetic field. While very few classical T Tauri star magnetic fields have been mapped in detail \citep{2014IAUS..302...25H}, application of the method has proven very successful including the case of AA Tau \citep{2010MNRAS.409.1347D} a known jet source. The magnetic field of AA Tau is shown in Figure \ref{AATauMag}.

In addition to analyzing the magnetic field of the young stars, it may also be possible with the next generation near-infrared (NIR) spectro-polarimeters, such as SPIRou \citep{2018haex.bookE.107D} to detect spectral features from the innermost regions of the accretion disks of high mass accretion rate stars (Class I and Class II). This could allow us to map both the $``$normal$"$ disk poloidal and toroidal magnetic fields for the first time in an analogous way to what has been achieved for a CTTS disk in the more excited FU Ori state \citep{2005Natur.438..466D}. A particular advantage of spectro-polarimeters like SPIRou is that not only do they operate in the NIR where disk emission can come from but also Zeeman splitting increases with $\lambda^{\rm 2}$, all else being equal, and this makes it easier to measure. Note also that this region of the disk is not accessible to ALMA. In addition, as SPIRou can measure radial velocity with exquisite precision it should be possible from the data to detect the presence of hot Jupiters. 

\subsection{Feedback: Outflow Interaction with Molecular Cores and Beyond}

There is no doubt that the final mass of a young star can be influenced by a whole host of environmental factors. As pointed out earlier, stars are frequently born in multiple systems that are dynamically active. Interaction between the companions can, for example, truncate a star's growth at an early phase \citep{2018MNRAS.475.5618B}. Radiation pressure from the newly formed star itself, particularly in the high mass regime, can limit accretion \citep{2019MNRAS.486.5171S} as can photo-evaporation from external sources such as supernovae and neighboring OB-stars \citep{2019MNRAS.485.3895H}. Nevertheless it is thought that
outflows may be very important in limiting accretion and thus the final mass a star can attain. Clearly they drive away material from the core and the envelope once they start to operate. It is not immediately obvious however what fraction of the final mass is removed and deposited back into the parent cloud.  A related problem is the effect of outflows on the molecular cloud itself. It has long been known that the lifetime of a molecular cloud should be shorter than observed if the only force acting against gravity is thermal energy. Instead an additional force must be present. In some cases this might come directly from the magnetic fields threading through a cloud. We also know however from observations that supersonic and super-Alfv\'{e}nic turbulence are present although the latter may only exist in the denser regions \citep{2012MNRAS.420.1562H}. The presence of turbulence alone however does not resolve the issue as the timescale for turbulence to decay in a typical cloud is short: there must be some local source of turbulence. Can this be outflows from young stars?

Certainly outflows create cavities in their surroundings while at the same time ejecting momentum and energy into it. The net effect is to disperse the core \citep{2006ApJ...646.1070A} and ultimately to terminate accretion. How efficiently outflows can do this determines the final stellar mass and also the fraction of the core's mass that is converted into a young star, i.e.\ the ratio between the so-called Core Mass Function (CMF) and stellar IMF (e.g.\ \cite{2014ApJ...784...61O}). As a specific example, the core mass function of low mass, star forming cores in the Pipe Nebula, was measured by using infrared extinction of back ground stars \citep{alves2007}.  This showed that the IMF is the direct product of the dense core mass function and a uniform star formation efficiency of $30\pm10\%$.  As already pointed out, both functions have similar profiles (see, for example \cite{2018ApJ...853..160C})
the difference being the peak in the CMF is at roughly 3 times the mass of peak in the stellar IMF. Obviously if outflows typically remove 2/3 of the core mass, this could explain the observed relationship. A very good overview of the processes that can determine the relationship between the CMF and the IMF can be found in \cite{2014prpl.conf...53O}. 

In additon the expectation that outflows sweep away such a large fraction of the surrounding gas over their lifetime is supported by recent observations of the HH46/47 system using ALMA \citep{2016ApJ...832..158Z}. In carrying out such observations it is important to use optically thin tracers, e.g.\ $^{\rm 13}$CO, to ensure the motion of the densest gas is not overlooked \citep{2006ApJ...646.1070A}.  

Numerical simulations, for example by \cite{2017ApJ...847..104O}, of the collapse and evolution of isolated dense cores now include the effects of turbulence, radiative transfer, and outflow feedback. These show that outflows can drive and maintain turbulence in the core environment even when magnetic fields are initially strong. Moreover, the final efficiencies are 15\%-40\% in line with the observed values. 
Of course these are simulations at the individual star level: it should not be neglected that multiplicity can play a role in feedback not only in multiple systems but also on the larger cluster scale. In multiple systems, interaction between the components can lead to precession of the associated jets \citep{2009ApJ...698..184W} thus broadening their impact on their surroundings. Moreover many stars are born in clusters, for example in regions such as NGC\,1333 containing hundreds of stars and large numbers of criss-crossing outflows \citep{2008hsf1.book..346W}. This suggests we need to consider the ensemble to get a clearer picture of how outflows feedback into their environment. 
Whether magnetic fields or turbulence driven by stellar feedback ultimately control how molecular cloud material is converted into stars is still an open question \citep{li_nakamura2006, 2016JPhCS.719a2002F}. 

\section{THEORETICAL OVERVIEW}

The last five years have also seen remarkable progress and new developments in the theory and simulations of magnetized outflows. The emergence of new computational tools and codes has played a leading role in these advances.  These include time dependent radiation field and 
chemistry within MHD simulations  \citep{gressel2015}. On the horizon are new capabilities including zoom-in simulations in radiation MHD simulations using DISPATCH \citep{nordlund2018}.  Post-processing tools now available include Monte Carlo radiation transfer methods in RADMC3D.  The ability to compute expected polarization maps from  general MHD systems has recently appeared in the POLARIS code \citep{reissl2016} which has been applied to the polarimetry of outflows \citep{reissl2017}. 

\subsection{Drivers for Outflows} 

There are several sources for magnetized outflows that are launched during the course of star formation, involving centrifugal or magnetic pressure drives from diverse sources such as forming disks in collapsing molecular cloud cores, the coronae of well established disks, magnetospheric boundaries, or the protostars themselves.  Before getting into the details, we outline this general theory landscape.   

(i) Magnetic braking and early outflows:  As magnetized, pre-stellar cores begin to emerge from the filaments within turbulent clouds, magnetic braking begins to extract their angular momentum.   The models of isolated, rotating molecular cloud cores with subsonic line widths that figure so prominently in many early theoretical and numerical studies are overly simplified pictures. In reality, hydrodynamic turbulence always contains small regions of non-zero angular momentum which, fed with enough mass, can produce disks - there is no need to assume that any coherently rotating object was present. In magnetized clouds in this earliest stage, turbulent MHD processes play a significant role in diluting the net braking torques on such pre-stellar cores, allowing disks to begin to form.  The magnetic field geometry in these chaotic initial conditions likely become more ordered close to the disks as dissipative effects such as ambipolar diffusion take effect in higher density regions.  Finally, at some point, collapsing magnetized cores reach their centrifugal balance radius.   With the formation of this disk, or disk-like system, the pinching of field lines and/ or the accumulation of toroidal magnetic field results in the launch of the outflow.  These are likely to global encompassing the entire eary disk.   Class 0 ouflows likely have their origin in this event.

(ii) Magnetized disk winds: will arise naturally during the evolution of magnetized accretion disks in their post formation phases. This has been shown by many authors, both theoretically and by means of numerical simulations using a variety of computer codes. These originate from disks that are both highly conducting, or in modern treatments, even from disks for which non-ideal effects dominate the physics of how fields are coupled to the gas and dust.  Details of the launching mechanism likely involve a thermal component to the disk wind as it is launched from the upper layers of the disk or base of a disk corona. The large scale suppression of the MRI instability discovered in non-ideal MHD processes means that disk physics itself is not controlled so much by viscosity as it is by the outward transport of disk angular momentum by powerful winds. 

(iii) Stellar spin and accretion powered stellar winds:  The solution of the angular momentum problem for the protostar itself is another key aspect of star formation physics.  If stars were to accrete the angular momentum from the inner edge of a Keplerian disk, they would spin up to break up speeds within a few hundred thousand years (the spin up torque of gas accreting from the magnetopause boundary $r_{MP}$ onto the star is then $\dot J \sim \dot M_w \Omega_{Kep,MP} r_{MP}^2$  \citep{2005ApJ...632L.135M}. An early suggested solution for this dilemma proposed that magnetic field lines from the star, connect back into the disk such that flux penetrating the disk beyond the co-rotating orbit would effectively brake the stellar spin by depositing angular momentum back into the disk \citep{koenigl1991}.  This $``$disk locking$''$ picture has the problem that magnetic field lines would quickly shear out and reconnect, severing the connection between the star and disk.  Centrifugally driven winds from the protostars can carry off angular momentum that is deposited by accreting gas streams flowing along magnetospheric field lines that connect the disk's inner edge, to the star \citep{2005ApJ...632L.135M}.  This work has met with considerable success in being a part of the rotational history of young stars \citep{bouvier2014}. Earlier models of X-winds proposed that angular momentum from the disk inner edge is never accreted onto the star but is carried off in an X wind originating from the interface between the disk and the protostellar magnetosphere \citep{2000prpl.conf..789S}.    

\subsection{MHD Outflow Basics}

To understand why magnetized outflows are so powerful and efficient in angular momentum extraction, we recall basic concepts of MHD flows from rotating bodies laid out in many papers and reviews \citep{ KoniglPudritz2000, 2007prpl.conf..277P, 2010LNP...794..233S}.  In the case of axisymmetric systems, we consider a rotating body threaded by magnetic field lines, with the footpoint of a field line located at a radial distance $r_o$ from the rotation axis, along the rotor. The angular velocity of this footpoint at the base of the flow is $\Omega_o$.  The basic equations of stationary, axisymmetric MHD flows (eg. \cite{pelletierpudritz92}, \cite{bai2016})  tell us that the mass flux is conserved along streamlines of the outflow (from the continuity equation), as is the magnetic flux $\Phi$. That is, the mass loss rate per unit area, along streamlines passing through an annular section  flow of area dA, is $ d \dot M_w / dA = \rho v_p $ (where $v_p$ is the poloidal velocity along the field line) while the amount
of poloidal magnetic flux through this same annulus is $d \Phi = B _p dA$. 
This means that the ratio of these two fluxes along a field line - the mass loss rate per unit magnetic flux, otherwise known as the mass loading of the flow - is also preserved:  $k = d \dot M_w/ d \Phi = const $. 

The mass loading is determined by the physical conditions at the foot point of the flow, which is essentially at the slow magnetosonic point on each field line.  As an example, if we consider the case of accretion disks as our basic rotors, the early 2D axisymmetric disk wind simulations of \cite{OuyedNature, Ouyed1997} assumed that the flow originates at the base of a heated disk corona, whose heat source is not specified.   Recent work takes the additional step of explicitly including FUV flux from the host star that heats the disk surface, feeding a disk corona \citep{bai2016}.   In any event, one needs to have a physical theory of how magnetic flux is distributed across the disk, and how mass is loaded onto it, in order to determine the mass loading function $k=k(r_o)$ which is a function of the conditions at the footpoint $r_o$  (ie the slow magnetosonic point) of each field line on the rotor.   

No theory that we are aware of has determined what this function is from basic first principles and so this remains one of the lingering uncertainties in the theory of MHD disk winds.  Authors adopt some sort of "reasonable" scaling for the magnetic field distribution across the disk.  As a specific example, consider outflows from magnetized Keplerian accretion disks.   Suppose that the threading vertical field component is a power law $B_z \propto r_o^{\mu -1}$, the outflow speed at the footpoint scales with the local Kepler speed, and the base of the disk corona can be considered a polytrope (eg. $\gamma = 5/3$).  The mass load function then scales as $k \propto r_o^{-(\mu + 1)} $.  As an example, \cite{1982MNRAS.199..883B}  assume a self-similar model for disk winds which imposes a specific scaling for the field $B_{z, BP} \propto r_o^{-5/4}$, so that $\mu_{BP} = -1/4 $ and hence  $k_{BP} = r_o^{-3/4}$. The stable minimum energy jet structures investigated by \cite{pelletierpudritz92} have $k_{PP} \propto r_o^{-1/2}$ \citep{pudritz06}. In general the power law index will also determine the degree to which poloidal magnetic field lines open up away from the footpoints.  Recent global MHD models of outflows \citep{wangbai2018} assume that heating of the disk surface by stellar EUV photons provides thermal heating and provides thermal pressure that assists in driving an MHD disk wind.  Here too, the initial distribution of magnetic field is assumed to follow a powerlaw distribution whose field lines initial structure is controlled by the  choice of the power law exponent.  

It is worth emphasizing that the concept of centrifugal acceleration of the flow arises naturally  in the co-rotating frame of the footpoint of the flow (e.g., \cite{2010LNP...794..233S}).  There is no magnetic force driving the outflow in this frame that role being played by the centrifugal force.  The flow in this region can be visualized as $``$beads on a wire$"$.  This is a very useful way to analyze the dynamics as long as the flow co-rotates with the rotor.  This ends roughly at the Alfvén radius, where magnetic and inertial forces balance. The collimation of jets and outflows begins after the outflow passes through the Alfvén point on each field line.  The reason is that the field lines now become increasingly toroidal.  The degree of outflow collimation depends quite sensitively on the mass loading of field \citep{ouyed1999, anderson2005}.  This is because it is the inertia of the material along the field line that causes the rotating field line to fall back from the rate of rotation of the footpoint, creating the toroidal field component.  This is much like the behaviour of a mass loaded whip.  The solution of the induction equation specifically shows that the toroidal field $B_{\phi}$ in such an outflow depends on the mass loading.   Since collimation of such rotationally driven outflows depends on the radial hoop stress exerted by the tension of the toroidal field line, mass loading must then control collimation properties of such winds.  Specifically, magnetic fields in disks with power laws  $\mu > -1/2 $ will collimate to cylinders, while lower values will lead to wide-angle outflows that are parabolic at infinity \citep{pudritz06}.  

As already noted, a central aspect of magnetized outflows is that they efficiently extract angular momentum from the rotor.  How does this happen?   Again, we first begin with ideal MHD since it allows us to see the basic physics quite simply.  The conservation of angular momentum equation for our magnetized rotating system says that the conserved angular momentum per unit mass $l$ of the fluid along a field line consists of two parts, the bulk rotation of the fluid itself $r v_{\phi}$,  and a component that is carried by the twisted mass -loaded field line, giving $l(r_o) = r v_{\phi} - ( r B_{\phi} / 4 \pi k) $. For any theory of stationary winds, solutions are obtained by requiring that they flow smoothly through a critical point on each field line (and for 2D flows, a critical line).  Critical points are positions at which the outflow speed is equal to a signal speed in the flow.   For purely hydrodynamic flows, the critical point occurs where the outflow speed equals the sound speed in the gas, $v_{p, hydro} = c_s$.  For MHD flows, there are three MHD waves: the slow and fast magnetosonic waves (compressive modes) and the transverse Alfvén wave.  The value of $ l$ on any field line is set by the condition that $v_p = V_A$, the Alfvén speed. If $r_A$ is the radial distance of the field line from the outflow axis where this condition is met, then $l= r_A^2 \Omega_o = (r_A/r_o)^2 l_o$  where $l_o = \Omega_o r_o^2$ is the specific angular momentum of the material at the footpoint of the flow. 

This result reveals why outflows are so efficient.  Each wind particle carries of a factor of $ (r_A / r_o)^2 $ more angular momentum per unit mass than its fellows rotating in the disk at the field line base.  Thus, if the typical lever arm is $r_A/r_o \simeq 3$, then each wind particle can carry the angular momentum of ten particles in the underlying rotor.  From this argument, it is also easy to see that in the limit that winds carry away all of the angular momentum of a disk that is accreting at the rate $\dot M_a $, the wind mass loss rate needed to drive this should simply scale as $ \dot M_a \simeq (r_A / r_o)^2 \dot M_w$.  Indeed, this result follows from disk wind theory outlined in the following subsection.  

The conservation of energy in ideal MHD, centrifugally driven winds is expressed by the Bernoulli equation for flow along the magnetic stream line.  One finds that the terminal speed of the outflow is greater than the escape speed from flow at the base of any field line, specifically $v_{\infty}  \simeq (r_A/r_o) v_{esc,o}$.  It is this energy relation that ensures that hydromagnetic outflows scale with the depth of the gravitational well of the rotor.   This is why star formation, from brown dwarfs to massive stars, is accompanied by jets and outflows. It is inevitably the efficient tapping of gravitational potential energy released during accretion by magnetized wind torques that is the main driver.  This also implies that because the terminal speeds of outflows on each field line scale with the Kepler speed of the footpoint, then disk winds will have a wide range of terminal speeds - the highest (hundreds of km per sec) in the interior parts of the jet or outflow, the lowest on largest outflow scales that can be supported on the disk (10 km per sec or less).  This is distinctly different than the prediction of X-wind models which really have only one velocity component originating from the innermost radius of the disk.   Observations generally support the former velocity structure (recall however, the SiO observations of HH212, which are consistent with an inner disk-wind or an X-wind origin; \citep{2017NatAs...1E.152L}).  

Another aspect of winds accelerated from disks is the launch angle at the footpoint of the flow.  An intriguing aspect of such outflows is that they can be launched for completely cold gas - one only needs the field line to be bent by an angle of $30^o$ from vertical \citep{1982MNRAS.199..883B}.  Of course, any additional heating of the outflow region near the surface of the disk will mean that radial thermal pressure gradients will increase this opening angle. As an example of this effect, polytropic models for the disk corona will have opening angles for field lines that increase with radial distance across the disk (eg. \cite{pudritz06}).  Vertical pressure gradients can also lead to better acceleration along the field lines.  Detailed heating models of disks by FUV irradiation from the star \citep{bai2016} will have their own prescriptions of field line opening.  

Early numerical simulations of jets focused on initial axisymmetric setups, often focused on the launch region near to the disk.  However, fully 3D simulations are needed to understand the complexity, stability, and asymmetries that can arise in jet dynamics in both the launch region and far from the disk.  

\begin{figure}[h!]
\begin{center}
\includegraphics[width=15cm]{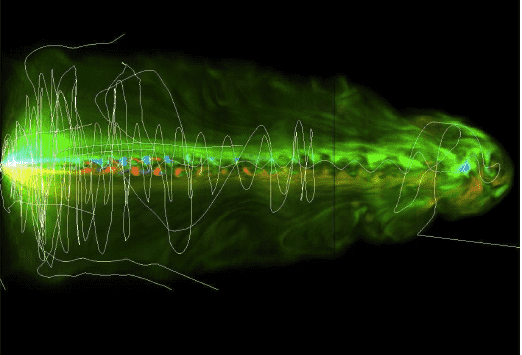}
\end{center}
\caption{3D MHD simulations of a disk wind showing the toroidal velocity and magnetic field lines for a disk wind produced in the \citep{pelletierpudritz92} model. The outflow source is the base of the corona of a Keplerian disk, shown on the left of the Figure. The simulations are run on a grid extending out to 90 au along the jet axis, and 27 au on either side of the jet axis. Brightness indicates higher radiation intensity from shock-produced, forbidden line emission in the jet. The red and blue colours of the knots indicate red and blue shift due to jet rotation.  Note also the recollimation of the jet half way down the flow.   From  \cite{staff2015} reproduced with permission \textcircled{c} OUP.  }\label{PP92}
\end{figure}

In Figure \ref{PP92} we show a global, high resolution 3D simulation, (using ZEUSMP- \cite{norman2000}) of a collimating disk wind whose source is in the corona of a Keplerian disk threaded with magnetic field lines \citep{staff2015}.  This figure shows the toroidal velocity and magnetic field lines in a jet launched from the corona of a Keplerian disk.  The magnetic geometry on the disk (and hence mass loading) follows that of \cite{pelletierpudritz92}. The jet is heated by shocks throughout its volume and the densities and temperature profiles are used to compute forbidden line emission in various lines in [CI IV], [S II],
and Mg II transitions, whose critical densities are $n \approx 1, 10^4, 10^5 $ cm$^{-3}$ respectively. The red and blue colours indicate Doppler shifts of rotating knots in the flow.  The jet field has a strong poloidal field along the axis of the jet which acts as a "backbone" that helps preserve the jet against instabilities.  Note that the toroidal field wraps around the jet, especially in the outer regions, giving rise to a highly collimated jet.  This as noted, agrees with observations of field structures seen in discussed in the observations of H80-81 (see Figure 4).  Other simulations with a less steep gradient of disk magnetic field did not collimate so well - as discussed above.

\subsection{Early Phase: Gravitational Collapse, Outflows and Disk Formation}

A problem of critical importance for stellar birth is the formation and evolution of protostellar
disks and the concomitant launch of outflows.  These are highly
nonlinear phenomena whose complexity arises from the interplay of gravity, turbulence, angular momentum, magnetic fields,
radiation, thermodynamics and even chemical evolution. Despite this, the appearance of disks and outflows in the earliest stages of star formation appears to be universal and this poses several important challenges for theory.  

\subsubsection{Idealized models}

The first theoretical treatments of star formation in magnetized clouds focused on magnetic braking of idealized, uniformly
rotating gas spheres.  This is enabled by a flux of a special wave: the torsional Alfven wave - first discussed in mathematical papers by \cite{1974Ap&SS..27..167G} and \cite{1979MNRAS.187..311G}.  Subsequently analytic solutions in ideal MHD were found by  \cite{1980ApJ...237..877M}.  The idea is simple to visualize.  Consider the formation of a condensation that is threaded with a magnetic field in a medium of non-zero angular momentum.  The ongoing condensation will generate a shear that twists the magnetic field, creating a toroidal field component.  The resulting magnetic torque on the condensation, extracts its angular momentum which is transported away by a flux of torsional Alfvén waves. The angular momentum is transferred to the surrounding medium which is gradually brought into co-rotation with the slowing rotor.   This braking picture suggests modest stirring of the environment around the forming star by a flux of these torsional waves. 
Given that the early stages of star formation are observed to be accompanied by vigorous outflows in the class 0 phase, this suggests that the braking phase occurs earlier, and sets the conditions for the first stages of disk formation. 

This simple model, however, was later realized to create a problem for disk formation.  Recent simulations have shown that braking in this picture is so efficient that disks  fail to form.  This has been dubbed as the magnetic braking catastrophe.  How does this arise?   
First we recall that the exact condition for collapse to occur in the first place is that the ratio of the cloud's gravitational to magnetic energy densities exceed unity.   The condition may be rephrased.  The ratio of these energies can be written in terms of the mass to flux ratio of the cloud -  which is equivalent to the ratio of the cloud's column density N to its field strength: $ (M / \Phi) \propto ( N / B) $ . The critical condition is that the ratio $\lambda $ of the cloud's mass to flux,  to the critical mass to flux for stability, exceeds unit:  $\lambda > 1$. The critical value $(N / \ B)_{crit} = 2.6  $ must be exceeded (ie. becoming supercritical) for collapse to occur \citep{HeilesTroland2005}.   \cite{LiMckee2015} performed MHD simulations of magnetized turbulent clumps and compared these to a variety of Zeeman and other observations and found that the typical molecular clump is supercritical with $\lambda \sim 3$. Thus, there is now a substantial body of observations confirmed by simulations, that places good limits on the origin and initial strength of magnetic fields in star forming cores - and therefore on the field that is available to drive outflows.  

A substantial body of work on collapse calculations focuses on the consequences of the Mestel picture of 
idealized isolated, spherical magnetized clouds, wherein the MHD is treated as either ideal or non-ideal \citep{2016A&A...587A..32M}. One of the key processes that sets in on small scales of disk formation at higher gas density is non-ideal MHD.   
Magnetic fields will decouple from the gas at high densities, and this, it was hoped, would lead to substantial decoupling of the field from the collapsing gas.  
Non-ideal MHD processes such as ambipolar diffusion
and Ohmic dissipation, which depend on how gas is ionized, and on grain sizes and charges at high densities, could play a role for the formation of
protostellar disks.  This has been investigated quite extensively by a number of groups.  

As an example,  \cite{mellon2008} showed that braking can prevent the build-up of disks during the early phase of star formation for relatively week fields; $\lambda \leq 10$.  Subsequent work established that even weakening the initial field by the process of ambipolar diffusion had little effect on this result \citep{mellon2009}.
The basic point is that by the time densities at which AD becomes effective are reached, magnetic braking is still so effective that only small disks may form.  Normal ambipolar diffusion and Ohmic dissipation are ineffective \citep{LiKrasnopolsky2011}.  High spatial resolution RMHD simulations carried out by \cite{Tomida2015} on the other hand, find that nonideal MHD
effects can resolve the magnetic braking catastrophe in the
early phase of star formation - before a protostellar core is
formed.  These simulations (using different inner boundary conditions, and refraining from the use of a sink particle) find the formation of a small rotationally supported ($\approx$5\,au) disk in the early phase. It is expected but not shown that the disk will grow later as larger angular
momenta accrete.  

One way out of this impasse for the classical model setup was suggested by \cite{hennebelle2009} who showed that by offsetting the magnetic and rotation axes of such rotors by even small angles of $\simeq 10^{o}$ centrifugally supported disks could form.  They found that centrifugally supported disks  cannot form for values of $\lambda \le 3$  when the magnetic field and the rotation axis are perpendicular, and smaller than about $\lambda \simeq 5-10$ under perfect alignment.  However, based on realistic
distributions of magnetic field strengths and misalignment angles,
\cite{2013ApJ...767L..11K} showed that this would lead
to a Keplerian disk fraction in the Class 0 stage of only 10 to at
most 50 per cent, noticeably below the observed fraction
of discs around Class I/II objects. 

\subsubsection{Realistic initial conditions: turbulence}

The magnetic braking catastrophe is really only a problem for the model of highly idealized initial conditions that it presupposes.   With the benefit of Herschel and ALMA observations, we now have a far better picture. Molecular clouds are characterized by supersonic velocity dispersion, attributed to supersonic turbulence \citep{Larson1981}. Herschel observations show that stars typically form in filaments \citep{2010A&A...518L.102A, 
	2014prpl.conf...27A}.  This likely arises from some combination of gravitational fragmentation of filaments into discrete cores, or as members of small groups or clusters at the intersection of filament systems.   However, examples of the theorist's dream model - fairly isolated, nearly spherical molecular cloud cores that are virtually identical to gravitationally unstable Bonner -Ebert spheres - do exist \citep{2001Natur.409..159A}!  Magnetic field observations show that star formation occurs in cores in which gravitational dominates magnetic energy \citep{crutcher2012}. 

The strength of magnetic fields, and hence the role of outflows in star formation, is rooted in how star forming gas becomes supercritical in the first place.  The intriguing point is that the cold neutral medium (CNM) out of which molecular clouds may ultimately form is highly subcritical \citep{HeilesTroland2005}. For the CNM, with measured total fields of the order $B_{CNM} = 6 \mu G $ (similar in value to the local Galactic component) these authors measure $\lambda = 0.42$ - deep into the subcritical, magnetically dominated regime.  Starting from this initial state of the ISM, an older approach to star formation in magnetized media supposed that the local mass-to-flux ratio is reduced by ambipolar diffusion until individual cloud cores become supercritical (see the review by \cite{lizano2015}).  This process is far too slow to be important in quiescent regions of molecular clouds (eg. star forming cores).  However, some analytical models \citep{zweibel2002} and numerical simulations \citep{nakamura_li2005} have suggested that this can be sped up in regions with supersonic turbulence.  

In supersonic turbulence however, intersections of supersonic shocks create filamentary systems (see the review by \cite{MaclowKlessen2004}).  
When magnetic fields are included, observations show that they tend to preserve their orientation at all scales that have been probed - from 100-pc scale down to sub-pc scale cloud cores. This suggests that both gravitational contraction and turbulent velocities should be anisotropic, due to the influence of dynamically important magnetic fields and this is in fact observed \citep{LiH-B2014}.
Simulations do show that because gas is free to flow along magnetic field lines - supersonic turbulence can compress gas flows along field lines \citep{padoan1999}.   While it has been claimed that gas motions in the denser cores are super Alfv\'{e}nic ( the density has increased with little change in the field \citep{padoan2018}), the observations suggest that the field is strong enough that motions are trans-Alfvénic.  The best studied low-mass cores of nearby star forming regions are typically subsonic. 

 Recent 3D global simulations of the initially subcritical ISM in a magnetized spiral galaxy show that long filamentary clouds form by the draining and compression of flows into long magnetic valleys of buckling field lines. This  onset of  such global Parker instabilities can create a system of kpc magnetized filamentary clouds of CNM .  Subsequent filamentary flow along these long HI structures gather sufficient gas to quickly 
 produce supercritical GMCs \citep{koertgen2018}.  The action of these global modes addresses the long standing question of how it is possible to gather enough gas from a collection region of the order of 1 kpc needed just to build GMCs (see review \cite{McKeeOstriker2007}). 

The origin of the initial angular
moment of pre-stellar cores is equally important. Given the fact the
molecular clouds are turbulent, it is questionable to assume that
angular momentum is coherently distributed. 
It is often assumed in analyzing the observations 
that cores have a uniform rotation and follow a
rigid-body rotation law. Their angular velocity $\Omega $  is deduced
from measurements of global velocity gradients in velocity maps \citep{MyersBenson1983,
Goodman1993}.
By performing numerical simulations of turbulent, magnetized clouds in 3D,
it has been found that the observations of the specific angular momentum 
of cores (which involve 2D projections) overestimate this quantity by 
a factor of approximately 10 \citep{dib2015}.  Hence, it is clear that protostellar disks are likely the result of {\it locally} generated angular momentum by turbulence \citep{JappsenKlessen2004}. In such situations, the link between the local angular momentum vector that will define the disk and the direction of the magnetic field at larger distances can be essentially random.   This, as it turns out, is the clue to resolving disk formation. 

The first simulations of disk formation in strongly magnetized, turbulent collapsing clouds were carried out using different numerical methods.  Simulations using 3D MHD Adaptive Mesh Refinement (FLASH) simulations on the scale of individual turbulent clumps were carried out by \cite{Seifried2012} and \cite{Seifried2013}. This work emphasized the inefficiency of magnetic torques in turbulent media. Another group focused on the role of turbulent reconnection of magnetic field lines in reducing the efficiency of magnetic braking \citep{santos-lima2012}.  The former approach showed that Keplerian disks of extent 50-150 AU can be formed in the collapse of turbulent clouds with the typical initial cloud magnetic fields; $\lambda = 2.6 $.    Disks form even in turbulent clumps with initial zero total angular momentum.  They arise as a consequences of the local angular momentum contained in the convergence of several filaments in the turbulent flow.  Disks formed for a range of initial core masses and turbulent settings, and for typical Mach numbers of the turbulence of $M= 5$, although disks formed in quite subsonic turbulence $M = 0.5$ as well \citep{Seifried2015}. In all cases, disks are formed by accretion out of discrete filaments and this is a fundamentally different picture than the collapse of idealized monoliths. The latter approach suggested that reconnection of tangled magnetic field lines plays a dominant role.  Although the interpretations of the results differ, it appears that turbulence can aid disk formation in magnetized media. 

Similar results have been found and extended by a number of different groups, using different numerical setups and codes.  \cite{Joos2013} find disk formation in magnetized turbulence and interpret this as a consequence of the misalignment effect that turbulence imposes on the forming disks. \cite{GrayMckee2018} ran simulations with similar initial conditions using their Orion 2 AMR, RMHD code, to the \cite{Seifried2013} setups - with a $300\, M_{\odot}$ initial clump, $\lambda = 2$, similar Mach number for turbulence, and most importantly, similar numerical resolution (1.2 AU). These simulations show, moreover, that rotationally supported disks don't form when the initial turbulent velocity field is aligned with the magnetic field.
 
 An analysis of the statistical properties of such disks formed in purely hydrodynamic simulations
of forming star clusters can be found in
\cite{2018MNRAS.475.5618B}. Although, this study does not include magnetic fields,
it confirms the general picture that initial velocity or density
fluctuations are sufficient to provide locally the necessary angular
momentum for the formation of protostellar disks. 

 Recent advances in computational techniques now allow, for the first time, multi-scale MHD simulations in turbulent Giant Molecular Cloud spanning scales from the cloud (40pc) down to a few AU scale on the disk scale  \citep{2017ApJ...846....7K}.  The transport of angular momentum in the disks can be followed.  The net inward mechanical transport is compensated for
mainly by an outward-directed magnetic transport, with a contribution from gravitational torques by spiral waves that is less than the magnetic transport.

\begin{figure}[h!]
\begin{center}
\includegraphics[width=12cm]{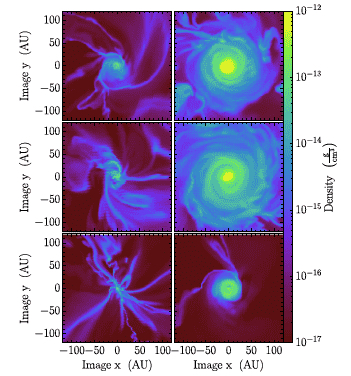}
\end{center}
\caption{Slices in the plane vertical to the mean angular momentum vector calculated for a sphere of 100 au at t=50 kyr. First row: sink 1 (left), sink 4
(right); second row: sink 5 (left), sink 6 (right); third row: sink 7 (left), sink 9
(right).  From  \cite{2017ApJ...846....7K} reproduced with permission \textcircled{c} AAS.  }\label{Kuffmeier_Sphere}
\end{figure}

\begin{figure}[h!]
\begin{center}
\includegraphics[width=12cm]{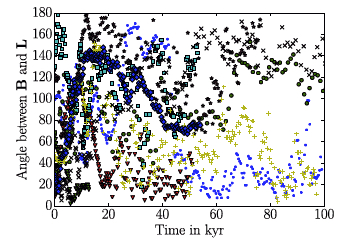}
\end{center}
\caption{Evolution of the angle between total angular momentum vector and total magnetic field direction within a sphere of 100 au around the eight different sinks. Note the high degree of random behaviour.  From  \cite{2017ApJ...846....7K} reproduced with permission \textcircled{c} AAS. }\label{Kuffmeier_Angle}
\end{figure}

 Figure \ref{Kuffmeier_Sphere} shows a selection of disks that form.  One sees that disks are often associated with filaments.  Because the most powerful parts
  of the jets are launched from the central regions of the disk, close
  to the (proto-)star, it is a
  numerical challenge to probe the efficiency of protostellar outflows
  in star cluster environments where spatial scale separation between
  the molecular cloud scale ($\sim 10\,\pc$) and the inner disc scale
  ($\sim 1\,\au$) spans more than six orders of magnitude. Attempts
  to probe self-consistently driven outflows out of turbulent cloud
  cores have been made.  Those simulations don't yet probe the innermost disk scales where the most powerful parts of the jet are launched, but the results are most encouraging.
  
Figure \ref{Kuffmeier_Angle} shows the evolution of the angle between the total angular momentum vector and the total magnetic field direction at 100\,au around 8 different sink particles (stars) in the simulation.  One sees that this distribution is essentially random, which is in good agreement with the observations.

 \subsubsection{Non-ideal MHD: effects of grain evolution} 
 
  As already noted,  for disks to form in ordered, laminar collapse, something must greatly accelerate the normal ambipolar diffusion time scale to reduce the field while gas is still at relatively low density.  This may be possible when one considers the fate of small dust grains in magnetized contracting cores. 

The large population of very small dust grains (VSGs) of order 10 - a few 100Å dominates the coupling of magnetic field to gas at densities $n > 10^{10}$ cm$^{-3}$.  The removal of this dust can lead to an increase of the ambipolar diffusion rate of the magnetic field out of the collapsing gas by two orders of magnitude. This allows 50-100 AU disks to form in well ordered collapses \citep{2018MNRAS.473.4868Z}. The depletion of VSGs arises as a natural consequence of their accretion of mantles (akin to molecular freeze-out)  or 
coagulation. Grain coagulation has been shown to remove small grains $ \le 0.1 \mu$m) within a few  yr \citep{Ossenkopf1993, Hirashita2012}.  The advantage of ambipolar diffusion driven by the loss of VSGs is that it occurs in the envelope of the forming disk which allows the formation of a large scale disk by avoiding catastrophic braking.  3D simulations of this picture show the presence of rings on scales of $ \sim 40$ AU.  

Since even subsonic turbulence allows for the formation of large disks in strong B fields, it is unclear at what stage this mechanism of grain removal in cores becomes important or competitive as subsonic conditions in cores are established.    

\subsubsection{The earliest outflows}

With the collapse of a turbulent overdense region of a cloud, the magnetic field configuration winds up creating an ever stronger toroidal field.  Angular momentum is extracted out of the region but since the collapse time scale becomes shorter than the time scale for the propagation of torsional Alfvén waves, the magnetized disk reaches its centrifugal balance radius $r_c$.  Because of braking, this radius will be considerably smaller than one would predict solely using simple angular momentum conservation arguments for the initial state \citep{Terebey1984}.  
As $r_c$ is reached, the winding of the field continues at a rate that is considerably faster than can be carried off by torsional wave braking.  Pressure confinement by the infall leads to the accumulation of "magnetic twist".  More precisely the building up of toroidal field creates an ever steepening vertical pressure gradient $ \partial / \partial_z (B^2_{\phi} / 8 \pi) $ in the outer reaches of the formaing disk out to the centrifugal radius.    It is this magnetic pressure gradient force that ultimately launches a tower flow.  The general idea of such flows was first derived theoretically by \citep{Lynden-Bell2003}.  

The first numerical dynamical simulations of the early stages of disk formation and magneto-hydrodynamic launching of protostellar jets were performed soon after the theory of MHD disk winds had been proposed \citep{1983ApJ...274..677P}.  In these simulations, magnetically threaded disks initially out of rotational balance,  collapse inwards and launch outflows \citep{Shibata1985, Shibata1986}.    This early work showed that strong transient outflows resembling tower flows were launched by collapsing rotating disks.  Subsequent work included a magnetosphere around the central star, and showed that a polar jet would also emerge from the star \citep{Uchida1985}. 

Tower-like outflows were  discovered and interpreted in magnetized collapse simulations by  \cite{BanerjeePudritz2006}.  The analytic solutions predict that the velocity of such towers is comparable to the rotation speed at the base.  In the collapse case, this implies an outflow roughly comparable to the rotation speed of the disk at $r_c$.  \cite{BanerjeePudritz2006} showed that the inner regions of the disk, where Keplerian rotation is firmly established, would give rise to centrifugally driven winds.  Thus, theory and simulations predict that disk winds at $r_c$ and inwards accompany disk formation.  This result corroborates the ALMA observations of large scale disk winds, shown in Figure \ref{OriSourceI}.

Simulations of magnetized collapsing cores now also show that the formation of the hypothesized first hydrostatic core is accompanied by the launch of a magnetocentrifugal wind.  Recent observations have found several candidates for these objects \citep{Enoch2010, Pineda2011} and that these are accompanied by slow, well collimated outflows \citep{Dunham2011}. \cite{Price2012} have used an SPH MHD code to show that low-mass first cores produce tightly collimated jets (opening angles $ \le 10^0$) with speeds of $2-7$ km s$^{-1}$, consistent with some of the observed candidates. 

\subsection{Main Phase: Disk Evolution and Outflows}

The vertically averaged angular momentum equation that governs a disk undergoing a total stress $\bf \sigma $  is  \citep{PudritzNorman1986,bai2016}, \begin{equation}
 \dot M_a {\frac{d}{dr}} (r u_{\phi}) = {\frac{d}{dr}} ( 2 \pi r^2 < \sigma_{r,\phi} > )+ 2 \pi r^2 \sigma_{z,\phi} \vert_{-h}^{+h} 
\end{equation} 
\noindent where the accretion rate is $ \dot M_a = 2 \pi r  \Sigma v_r$ for a radial inflow speed of the gas $v_r$, and the angle brackets in the first term indicate taking the vertical average of the torque by integrating over z.  The total stress has contributions from both turbulence, and the Maxwell stress of threading magnetic fields.  The first term on the right hand side denotes angular momentum flow in the radial direction, while the second term is angular momentum flow out in the vertical direction due to wind torques.  In the case of shear turbulence, the stress is  the average of the turbulent fluctuations, 
$ \sigma_{ r, \phi} =  -  \rho  \delta  v_r  \delta v_{\phi}   $.
In the presence of  a toroidal magnetic field $B_{\phi}$ in the disk, a radial field $B_r$ can also contribute to flow in the radial direction through  the Maxwell stress component;
$\sigma_{r, \phi} =B_r B_{\phi}  $.   This possibility arises naturally in diffusive models due to field line dragging in the accretion flow.  It also appears in recent models of non-ideal MHD wherein the Hall effect can produce an instability leading to a radial field component  \citep{2014ApJ...796...31B, Bethune2016, 2017MNRAS.472.1565M}. A threading vertical component of the field $B_z$  however, exerts a torque on the disk with $\sigma_{z, \phi} = B_z  B_{\phi}  $ leading to
 an MHD disk wind, which is central to the action of the ubiquitous jets and outflows that accompany the formation of all young stars, regardless of their mass \citep{2014prpl.conf..451F, 2007prpl.conf..231R, 2007prpl.conf..277P}.  

 Physical models of accretion disks have focused heavily on the assumption that angular momentum is transported  by turbulent viscosity, first addressed in the seminal papers by \citep{SS1973, LBP1974}.
 Here, the turbulence is assumed to arise from the shearing Keplerian flow and takes the form $ \sigma_{r, \phi} = \nu \Sigma r d \Omega / dr $. The effective viscosity of the disk $\nu$ can then be shown to scale with the disk scale height as $ \nu = \alpha c_s h$ with the famous $ \alpha $ parameter.    
Steady state disks then have a radial accretion rate $ \dot M_a $ , which, away from the inner boundary of the disk can be written as
 \begin{equation}
 \dot M_a = 3 \pi \nu \Sigma = const 
 \end{equation}

In order to drive an accretion flow at the rate observed to fall onto T-Tauris stars, $\alpha \simeq 10^{-2} - 10^{-3} $. The angular momentum is carried out radially leading to the slow, outward radial spreading of the disk from its initial state.
The high column densities of protoplanetary disks prevent much radiation from penetrating the disk, leaving it poorly ionized at the midplane.  Thus non-ideal MHD processes such as Ohmic losses, ambipolar diffusion, and the Hall effect, all take their toll on the coupling of magnetic fields to gas.   
Each process dominates in a different vertical layer of the disk: Ohmic diffusivity in the mid plane, the Hall effect in the middle layer, and ambipolar diffusion (AD) in the surface regions.  These diffusivities vary as a function of radius across the disk.  As an example, the Ohmic diffusivity depends on both the temperature and electron fraction, both of which change with disk radius (eg. \cite{cridland2016}).  
The damping of MRI instabilities occurs when the ratio of the growth rates to the damping rates predicted by 
these diffusivities is less than unity (e.g. review \cite{Turner2014}).  These are the so-called  Els$\ddot a$sser numbers for each effect: $ A_m =   v_A^2 / (\eta_A \Omega)$, 
$\Lambda_H=  v_A^2 / (\eta_H \Omega)$, and $\Lambda_O =  v_A^2 / (\eta \Omega)$.  At the disk midplane, where dust processes are important for building planets, Ohmic diffusivity generally dominates in the inner disk radii. However, ambipolar diffusion can start to dominate in the diffuse outer portions of the disk.  
 Early work \citep{Gammie1996} noted that whereas the MRI would be damped at the disk midplane by Ohmic dissipation,   MRI driven accretion could still occur in an “active layer” in the disk surface regions.  This idea known as the "dead zone" is an important region in the disk for many reasons, including issues of dust growth and planet formation.  However, the inclusion of AD shows that the MRI instability is damped even in these surface layers \citep{baistone2013}. What does this say about outflows?  

The  launch of outflows requires a treatment of the mass loading of the flow. This demands a sensitive treatment of the microphysics of the gas at the magnetosonic point.  Wind launch by a centrifugal mechanism requires that the field be sufficiently bent away from vertical.  Vigorous outflow requires that the field be reasonably strong as well – near to equipartition value with thermal pressure in some models – making it difficult to bend the field as it emerges from the disk surface layers \citep{wardlekoenigl1993, ferreirapelletier1995},   However AD effects allow bending to occur \citep{1996ApJ...465..855L}.  It was also noted that a strong vertical field would reduce the rotation of the disk to sub Keplerian rates too small to drive the wind \citep{ogilvie2012}.   Analytic treatments that included the full roster of non-ideal MHD effects seemed to suggest that MRI and centrifugal wind effects are mutually exclusive \citep{koenigl2010}.  Disk winds operate in weak MRI and field strengths near equipartition, while MRI requires weak fields at the midplane and would then not produce accretion rates high enough to match the observations.  
 
Detailed simulations of the physics of these processes have led to a qualitative leap in our understanding.  \cite{baistone2013} simulated vertically stratified local shearing patches of disks threaded by weak vertical field lines,  with a self consistent treatment of Ohmic and AD diffusivities based on detailed disk chemistry.   They found that although MRI quickly develops in the initial weak disk field,   the disk rapidly adjusts to a new equilibrium state in which the disk is laminar (the turbulence is almost completely damped) and a magnetocentrifugal outflow is formed.  The viscous stress parameter was only $\alpha \simeq 3\times 10^{-6}$  with viscous  transport restricted to a narrow FUV heated surface layer.  The angular momentum is carried off by the wind which drives disk accretion at rates sufficient to match the observations - but not so strong as to rapidly deplete the mass of the disk.   All of the disk accretion flow takes place in a current layer of thickness $ 0.3 h$ (where h is the disk scale height), at radial inflow speeds about $0.4 c_s$. Disk evolution in these regions is unlikely to depend on turbulent viscous forces, although very low level turbulence may still be excited by various hydrothermal instabilities \citep{flock2012}.

\begin{figure}[h!]
\begin{center}
\includegraphics[width=12cm]{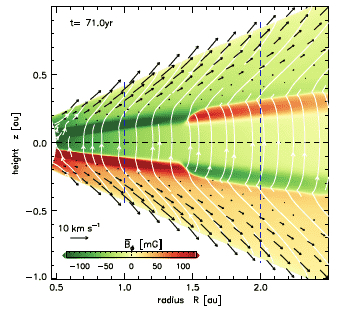}% This is a *.eps file
\end{center}
\caption{ Plot of the toroidal field (colour) with peak values of a few hundred mG. The projected magnetic field lines (white) and velocity vectors (black) are indicated as is the position of the wind base (dot–dash).  From \cite{gressel2015} reproduced with permission \textcircled{c} AAS.  } \label{Gressel}
\end{figure}

\begin{figure}[h!]
\begin{center}
\includegraphics[width=12cm]{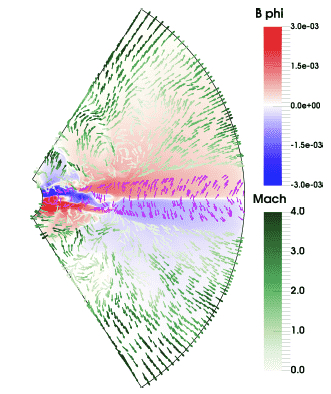}% This is a *.eps file
\end{center}
\caption{A 2D, global non-ideal simulation of a disk and outflow showing: the toroidal magnetic field in color background (blue to red), the poloidal velocity field in units of local sound speed (Mach number) - green arrows in the corona, and the orientation of the angular momentum flux caused by magnetic
stress (purple arrows in the disk).  From \cite{2017A&A...600A..75B} reproduced with permission \textcircled{c} ESO.  } \label{Bethune}
\end{figure}

Global simulations of non-ideal MHD effects in disk winds have been performed using different computational techniques. 
In Figs \ref{Gressel} and \ref{Bethune} we compare results of the simulation of these effects from two different research groups \citep{gressel2015, 2017A&A...600A..75B}.  
 The global disk simulations of \cite{gressel2015} include both Ohmic and AD terms and are shown in Figure \ref{Gressel}.   Their results are very similar to the shearing box simulations of \cite{baistone2013}.  In particular - the inclusion of AD completely quenches the turbulence in the disk surface layers: without AD effects, MRI instabilities develop that transport angular momentum in the surface layers.   In the full simulations, the field at the midplane is dominated by its poloidal component because the strength of the Ohmic dissipation effects prevents the appearance of currents that would create a toroidal field.  As one moves away from the midplane, the field bends radially outwards.   The reason for this is as follows (eg. \cite{KoniglPudritz2000}).  The ions are braked directly by the magnetic torque and so move with slightly sub-Keplerian velocity.  This is transmitted to the neutrals by the frictional drag (ambipolar diffusion), which slow down and lose their angular momentum to the field (the toroidal field increases). Thus the neutrals must  move radially inwards in the disk.  In so doing, the field is also dragged inwards, causing the field lines to begin to curve (in the R-z plane).  The magnetic tension resulting from this field line curvature then balances the drag force. 
 
 The direction of the toroidal field in the inner part of the disk (inside 1.5\,au) is in the same direction as that of field lines in the outflow. The poloidal field lines bend smoothly into the base of the outflow region.  At disk radii $ r \ge 1.5$\,au, the polarity of the toroidal field reverses.   Field lines now start to curve towards the star as the first steps in flow collimation start to take place.       The change in toroidal and radial field components takes place in a narrow current layer concentrated at $ z \simeq \pm 3.2 h$.  These layers are found to be stable.  Launch of the winds can be achieved with weak initial magnetic field strengths in the disk. 

In Figure \ref{Bethune}, the Hall term is also included within a full 3D global simulation of the disk \citep{2017A&A...600A..75B} although detailed chemistry and radiative transfer calculations are not included.    There are similarities with the \cite{gressel2015} results.  The toroidal field changes sign across the midplane.  It also changes sign at a disk radius ($2.5 r_o$), largely an effect of the inner disk boundary conditions.  An intriguing aspect of this model is that the disk can be accreting, or not, depending on the configuration of the large scale magnetic field.  In this Figure, the inner region of the disk has a flow of angular momentum from the disk surface towards its midplane – while the outer regions have angular momentum flow away from the midplane.   In general,  angular momentum transport towards the mid plane results not in disk accretion, but rather large scale meridonal circulation. The non-accreting case is characterized by a meridional circulation, with accretion layers at the disk surface and decretion in the midplane. In the accreting case, vigorous disk winds drive radial accretion flows.

One of the intriguing aspects of the Hall effect is that the direction of transport of the magnetic flux in disks depends on the polarity of the threading poloidal field component $\bf {B_p} $ with respect to the disk rotation axis.  If its direction is parallel to {\bf $ \Omega$ }, then flux transport 
 is inwards, and if anti-aligned, outwards \citep{2017ApJ...836...46B}.  Since the flux distribution affects the strength of the wind torques,
 these Hall effects could be significant for the physics of Type I migration (migration of low mass bodies that do not open gaps in the disk).    
 In all situations, it appears that disks do not support MRI turbulence out to distances of 10 AU for standard conditions.  This dead zone radius $r_{DZ}$ must evolve with time as the disk thins out.  
 It is particularly noteworthy that 10 AU scales corresponds to the planet forming region in most theoretical models of planet formation.

\subsection{Stellar spin and outflows} 

At the earliest stages of their evolution,  stars are in the process of contracting and are often accreting material with high specific angular momentum  from a
protostellar  disk.  It is surprising, therefore, that a large fraction of stars rotate more slowly than expected - approximately half of solar mass
stars with ages of less than a few million years rotate at
less than $10 \%$  of their breakup velocities - ( eg. \cite{Rebull2004}, \cite{Herbst2007}; \cite{Scholz2009}).  In the absence of a significant spin-down torque, most stars should spin at near breakup speeds.  Accretion disk signatures are more prevalent among slowly rotating pre main sequence stars, suggesting that accretion and rotation are connected. 

The idea that protostellar winds could drive protostellar spin evolution has been suggested by a number of authors \citep{Hartmann1982, Tout1992, Paatz1996, Ferreira2000}.  \cite{2005ApJ...632L.135M} showed that as long as the mass loss rate in magnetized stellar winds is high enough, then these would dominate over disk-star coupling spin-equilibrium mechanisms.  Specifically, they proposed that accretion onto the star would power a strong wind by the excitation of a large flux of Alfv\'en waves along the open field
lines.  They called this new class of stellar wind models - accretion powered stellar winds (APSW).  This wave excitation mechanism was further explored by \cite{Cranmer2008} and  \cite{Cranmer2009}, who showed that
the predictions for X-ray luminosities from the shocks in these models are in general agreement with existing observations.

Theoretical studies of accretion powered stellar winds (with mass loss rate $\dot M_w$, stellar dipole magnetic field $B_*$, and radius $R_*$), derived the scaling of the Alvén radius, expressed as the lever arm of such a wind; 
\begin{equation}
(r_A /  R_*) = K ( B_*^2 R_*^2 / \dot M_w v_{esc} )^m
\end{equation} 
where $K = 2.11$ and $ m = 0.223 $ are found by fits to numerical calculations \citep{Matt2008}.  We note that this general scaling relation also holds in the context of disk winds, where from \cite{pelletierpudritz92} one can deduce that $ m=1/3$.  

The APSW model was applied to the study of spins of young stars undergoing evolution down the Hayashi track - up to 3Myr in their evolution \citep{Matt2012}. The results were compared with X-wind and disk locking models.  Overall, APSWs explain the observed distribution
of young star spins in a similar way as the classical disk-locking
picture, while at the same time avoiding the problem of magnetic
field line opening it necessarily entails. When compared to the X-wind picture, it avoids  the
assumption of spin equilibrium and requirement of significant
flux trapping.  For the best fit to the  observations, the APSWs are predicted to have mass loss rates that at least $ 1 \%$ of the disk accretion rate.  

It is observationally challenging to clearly isolate an APSW from the fastest components of a disk wind originating at the innermost regions of the disk.  However, it may be possible to identify the source regions of the flow predicted by the coronal wave- heating picture that it relies upon.  

\subsection{ Outflow feedback: determining stellar masses and star formation efficiency} 

Given the efficiency of magnetized outflows in tapping gravitational potential energy release and the angular momentum of disks, how is the surrounding core affected? Is stellar mass determined by this outflow feedback? And  given that most stars form in clusters, do outflows control the star formation efficiency of clusters? 

 Most of the gravitational energy released during accretion originates from the inner disk.  Disk winds scale with the local escape speed at the base of the footpoint of the flow, and hence with the escape speed from the central star.   Since both low and high mass stars have approximately the same escape speeds, it follows that low mass stars are just as effective as high mass stars in their feedback effects \citep{MatznerMckee2000, KrumholzARAA2018}. This is unlike any other feedback mechanism (eg. radiation fields) whose effects are strongly dependent on stellar mass.  

The momentum per unit mass that can be delivered by these outflows is of the order 40 km s$^{-1} $ which if spherically distributed, would disrupt cores.  However outflows are well collimated and the interaction between outflows and the core material is mediated via a thin radiative shock in the \cite{MatznerMckee2000} model. This model also assumes that there is an escape polar angle $\theta_{esc} $ that separates the outflow from the region that accretes through the disk onto the star.  The most important aspect of the collimation of the flow is that the thrust is concentrated in a mass that is less than the spherical case by a factor $1/ln(2 \theta_o)$ where $\theta_o$ is the angular extent of the central outflow (typically $\theta_o \simeq 10^{-2}$). 

This model assumes that accretion onto the star continues through the disk as the disk wind carves out a wind cavity. The ultimate mass of the star, therefore, is determined not by the outflow, but by having a finite mass reservoir (the core) out of which accretion proceeds.  The prediction is that as long as low mass star formation dominates, true for clumps and clusters of $\le 100 M_{\odot}$, that the efficiency of star formation $\epsilon = M_*/ M_{clump,core} = (v_{esc} / v_w) ln (2 \theta_o ) / f_w \simeq 0.3-0.5 $, where $f_w$ is the ratio of the wind to star mass.  This provides a theoretical explanation for the difference between the core mass function for the gas out of which stars form, and the initial mass function of stars, which are well known to differ by a factor of three in overall mass (eg. review \cite{2014prpl.conf...27A}. 

Recent numerical simulations allow a more comprehensive look at the complexities of this problem.  High resolution computations down to the stellar surface have been done \citep{Tomida2015}, but the small time steps required limit these to only a small span of time (a few years). To overcome this, a number of simulations resort to building subgrid models that effectively prescribe the physics of an MHD outflow whose source region remains unresolved 
\citep{federrath2014}. As already noted, \cite{2017ApJ...847..104O} find that the inclusion of real MHD effects in cores reduces the star formation efficiency - lower mass to flux ratios lead to a decrease in the star formation efficiency; $\epsilon $ drops from 0.4 to 0.15 as $\lambda$ is decreased from hydro $\lambda \sim \infty$ to $\lambda = 1.5 $.  These simulations also find that ratio
of launched outflow to the total (combined
launched and entrained) outflow mass is 1:4.  

The question of whether or not jets can drive turbulence has been addressed by several different methods and model setups.  Maintaining turbulence can help delay gravitational collapse.  The basic energetics of jets provides some interesting insights on this feedback \citep{banerjee2007}. The lifetime of a jet can be estimated as $\tau_{jet} \simeq L / v_{jet} $ which for speeds of $\sim 300$ km s$^{-1}$ and a typical length scale of $L \sim 3 $ pc is about $10^4$ yrs - the typical duration of the early phase of star formation (class 0).  The total mechanical luminosity of a jet $L = \dot M_{jet} v_{jet}^2 / 2 $ is of the order $3 \times 10^{33} $ ergs s$^{-1}$ for a jet mass loss rate of $\simeq 10^{-8} M_{\odot} yr^{-1}$. Thus the total energy supplied to a cloud by a jet is $E_{jet} \simeq L_{jet} \tau_{jet}$, which is of the order $10^{44}$ ergs.  For a cluster forming clump of $10^3 M_{\odot}$, the total turbulent energy given a typical velocity dispersion of $\simeq 1$ km s$^{-1}$ (Mach 5 turbulence) is $E_{turb} \simeq 10^{46}$ ergs.  Thus, if $f$ represents the coupling factor of converting outflow power into turbulence, the number of outflows needed to maintain turbulence in such a region is $N \sim 100 / f $.  The lifetime of the turbulence scales with the crossing time of the region $L / v_{rms}$ and is of the order a few $10^6$ years.  Thus, if the coupling is reasonably strong, jets can power clump turbulence and this has an effect on star formation efficiency.

How efficiently does jet power convert to tubulence?  \cite{banerjee2007} found in 2 and 3D simulations that jets injected into quiescent cores do not develop instabilities (Kelvin-Helmholtz) and turbulence.  The basic point here is that such instabilities are best excited for subsonic velocities.  Jets, under these conditions, therefore have only very limited ability to drive supersonic turbulence.  Other studies include multiple, highly collimated flows injected into an initially quiescent \citep{carroll2009} or  turbulent medium \citep{2017ApJ...847..104O}) and these find that turbulence is excited.  It is unclear what the reason is for such different results. Simulations of turbulence generated by jets in an initially turbulent clump suggest that it may be protostellar jet generated turbulence that sets the conditions for turbulent fragmentation into the IMF of stars in the cluster \citep{li_nakamura2006}.

\subsubsection{Massive star formation and feedback}

Massive stars start to appear in clumps that exceed $100 M_{\odot}$, and the nature of feedback related to radiation fields (photoionization and radiation pressure), massive stellar winds, and supernovae feedback are dependent on stellar mass (eg. reviews, \cite{McKeeOstriker2007, Tan2014}).
\cite{Fall2010} adopted semi-analytic arguments to compare the general effectiveness of protostellar outflows and other feedback mechanisms.  Outflows are only effective at feedback for clumps with escape speeds below $7$ km s$^{-1}$.  In the context of highly idealized, spherical cloud models, radiation pressure is the dominant feedback process that controls star formation efficiency in massive star clusters ($\ge 10^4 M_{\odot}$).  For lower clump masses, especially below $10^3 M_{\odot}$, protostellar outflows can play a significant role.

\begin{figure}[h!]
\begin{center}
\includegraphics[width=15cm]{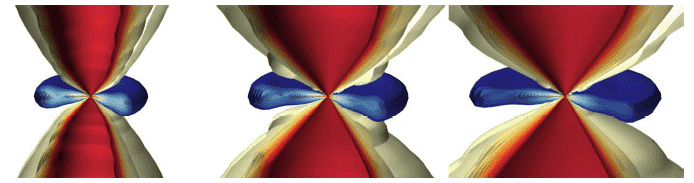}% This is a *.eps file
\end{center}
\caption{Massive star formation and radiation feedback from jets and radiation fields.  From left to right: accretion disk and protostellar outflow at times t = 20, 30, and 40 kyr of evolution. From \cite{kuiper_hosokawa2018} reproduced with permission \textcircled{c} ESO.  } \label{Kuiper1}
\end{figure}

\begin{figure}[h!]
\begin{center}
\includegraphics[width=12cm]{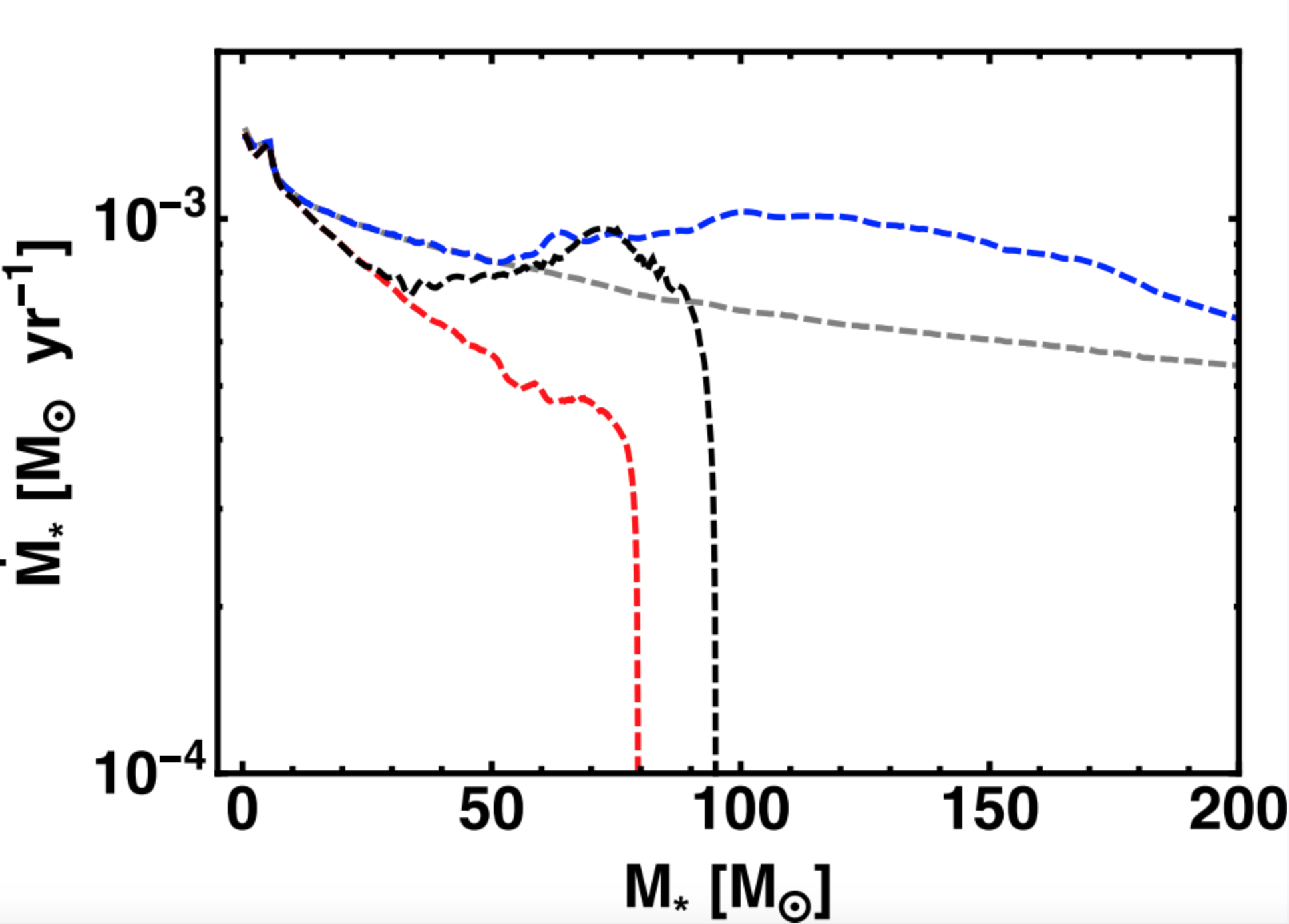}% This is a *.eps file
\end{center}
\caption{The role of radiation and protostellar jet feedback in the formation of massive stars. Gray  line is for outflow feedback only; blue line, outflow +  photoionization feedback; red line, ouflow + radiation forces; black line, outflow + radiation forces + photoionization.  From \cite{kuiper_hosokawa2018} reproduced with permission \textcircled{c} ESO.  } \label{Kuiper2}
\end{figure}

This leads to the question of whether magnetized disk winds play a role in the formation and feedback processes of massive stars. There is no consensus
on the precise upper mass limit for stars, but the evidence to date suggests 150–300\,$M_{\odot}$ \citep{Figer2005, Crowther2010}.  Massive star formation likely involves the accretion of material from a massive disk which for a short time of 10-20  kyr,  may be comparable to the mass of the star forming at its centre.   Because of accretion flows through massive disks, the limitation on stellar mass of $\simeq 40\,M_{\odot}$ set by the Eddington limit for spherical collapse is avoided by having radiation emitted through a radiatively driven cavity \citep{Yorke1999}. This anisotropic distribution of radiation due to the optical depth effects of the underlying flaring accretion disk is known as the "flashlight effect".  The initial conditions for massive star formation resemble, to some degree, those for cluster formation.   Many simulations often begin with an isolated clump mass of $\simeq 10^3 M_{\odot}$.  

Early simulations of photoionization effects from massive stars employed ray tracing methods within the FLASH AMR code \citep{Peters2012}.  These simulations analyzed the role of photoionization and HII regions in driving observed molecular outflows from massive stars. Feedback from
ionising radiation was not sufficient to drive the observed massive CO outflows around massive stars. This suggests that magnetically driven outflows are important even around massive proto stars. 

High spatial resolution, RHD simulations show that radiation pressure from the forming massive star pushes a bubble into the collapsing gas, while disk accretion continues through the disk \citep{Krumholz2009,  Kuiper2010, 2016ApJ...823...28K, Rosen2016}.   The details of the RHD simulations matter.  \cite{Krumholz2009} used an Adaptive Mesh Refinement code (Orion) with radiative transfer handled only through flux limited diffusion, and found that the continued mass accretion onto the star proceeded by infall of cold gas produced by Rayleigh-Taylor "fingers" on an unstable bubble wall.  \cite{Kuiper2010} used a high resolution, spherical fixed grid with a ray trace RHD method and found that such fall back did not occur, and that strong disk accretion flow built the star.  The maximum stellar mass attained was $210 M_{\odot}$.  \cite{2016ApJ...823...28K} combined an AMR approach with a hybrid approach to radiation transfer - combining ray tracing techniques with flux limited diffusion.  These simulations found that indeed, the main mode of mass transfer to the star was by strong accretion flow, mediated by spiral waves excited in the self gravitating disk.  The bottom line in these latter two studies is that by using only RHD, the mass of a star is limited only by the size of the mass reservoir.  However, this group of studies did not include magnetic fields. 

Because MHD disk winds are launched at the moment of disk formation, massive star forming disks should have already started to carve out a cavity before the radiation fields become important. In this regard, it is interesting to note that the best case for rotating disk winds is for the massive protostar in Orion as shown in Figure 1.  Radiation, once it does become important, will be able to escape through this outflow cavity - making the issue of the details of radiation blown bubbles a somewhat moot point \citep{Krumholz2009}.  It is important therefore to have combined RMHD simulations of massive star formation.  Will MHD outflows make it easier to form more massive stars?

Recent simulations \citep{kuiper_hosokawa2018} include radiation forces from direct and dust-reprocessed radiation as well as photoionization.  The effects of a magnetized disk wind were included: the outflow was injected into the
computational domain at $t = 4 kyr$  with an ejection-to-accretion efficiency of $10 \% $, and a velocity of three times the local escape velocity with respect to the current protostellar mass.  The evolution of the protostar was followed using evolutionary tracks by \cite{HosokawaOmukai2009}. 

Figure 11 shows the resulting structure and evolution of the outflow and disk.  The overpressure of the HII region fills the outflow cavity (left frame).  This pressure acts on the surface of the disk below the cavity, serving to compress it and in the process, opening up the cavity even more.  This enhanced pressure due to photoionization leads to the pile up of additional mass on the disk (middle frame) which shadows the disk beyond.  This shadowing enables an  enhanced accretion rates through the disk (the ''scissor effect''). Other than this effect, photoionization does not seem to play a big role in this highly optical thick disks.  

Figure 12 shows the accretion history and final masses of stars undergoing various combinations of feedback: outflow alone (gray), outflow + photoionzation (blue), outflow + radiation forces, and lastly (red) all 4 feedback effects (black).  We see that the magnetized outflows do not stop the accretion from the very large $1000 M_{\odot}$ mass reservoir.  Adding photoionization to the picture also does not shut off the infall.  The addition of radiation pressure to the outflow, however, does lead to the truncation of accretion even from an effectively infinite mass reservoir (red and black curves).  The final masses are 79 and 95 $M_{\odot}$ for the red and black curves, respectively.  The photoionization scissor effect, present in the latter case, helps increase the accretion rate and final mass.  In addition to the protostellar
outflow, radiation forces are the dominant broadening mechanism of the bipolar region.   
These results suggest that, like low mass star formation, disk winds by themselves create feedback, but probably do not determine the masses of stars.  

\begin{figure}[h!]
\begin{center}
\includegraphics[width=15cm]{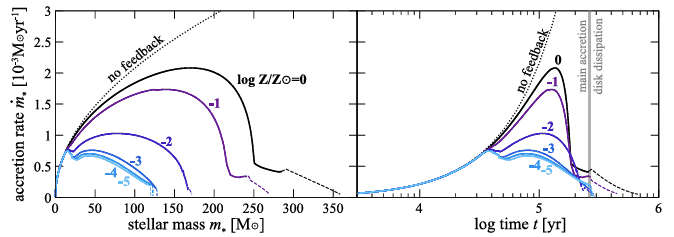}% This is a *.eps file
\end{center}
\caption{The role of magnetized outflows in massive star formation as a function of metallicity.
Accretion histories as functions of protostellar mass, m$_{\star}$, (left) and time, t, (right) for stars forming from cores with initial masses of 
M$_{\rm c}$=1000\,M$_{\odot}$ and
embedded in clump environments with $\Sigma_{cl}$ = 1 g cm$^{-2}$. 
Results for metallicities log Z/Z$_{\odot}$ = -5, -4, -3, -2, -1, and 0 are shown as labeled. For each line, the
solid part represents the main accretion phase and the dashed part is the disk dissipation phase (the gray vertical line in the right panel indicates the transition time).
The black dotted lines show the no-feedback case. The accretion rate is lower at lower metallicity due to stronger total feedback.  
From \cite{tanaka_tan2018} reproduced with permission \textcircled{c} AAS.  } \label{Tanaka}
\end{figure}

We also show a study that comes to different conclusions - as is shown in Figure 13 \citep{tanaka_tan2018}. In this work, a semi-analytic treatment of radiative feedback and MHD disk winds is adopted.  The latter input generalizes the \cite{MatznerMckee2000} model. An important new addition to the analysis is the dependence of disk accretion and final stellar masses on the metallicity.  Radiation pressure is found to play a minor role in the feedback mechanism - MHD disk winds are the key player providing $\ge 90 \%$ of the outflow momentum. The novel insight in this work is that as the metallicity decreases, photoevaporation becomes stronger. This reduces the SFE because dust attenuation of ionizing photons is inefficient.  In this analysis, there does not appear to be a firm upper limit to stellar mass.

The somewhat different conclusions that these studies reach likely arise from differences in their underlying modeling. The first is the different level of detail in the radiative transfer calculations, full RHD vs semi-analytic theory.   The second is that it may be useful to include the MHD wind explicitly. The subgrid modeling of the MHD disk wind itself, while capturing several of the key points, may still require more dynamic treatment.   That said, it is clear that radiation and MHD disk winds are both important in massive star formation, and this is a tantalizing open problem.

\begin{figure}[h!]
\begin{center}
\includegraphics[width=15cm]{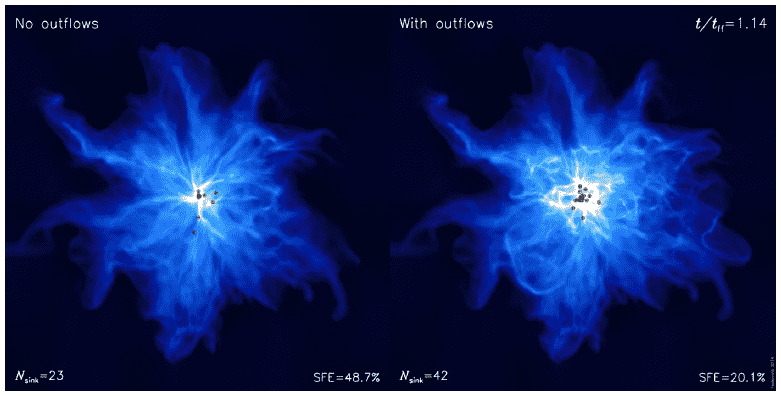}% This is a *.eps file
\end{center}
\caption{Effects of feedback from protostellar outflows on star and cluster formation. The left frame shows the results of cluster formation without outflow feedback, and the right when feedback is included.  Note the strong effect on the SFE and number of stars produced.   From \cite{federrath2014} reproduced with permission \textcircled{c} AAS. } \label{Federrath}
\end{figure}

Finally, we return to the question of the role of outflow feedback in cluster formation, as investigated by numerical simulations.  \cite{federrath2014} applied  the MHD disk wind, sub grid model to turbulent, magnetized star cluster formation. The subgrid model features a disk wind opening angle of $30^o$ - the minimum angle that gives rise to outflow for cold initial conditions \citep{1982MNRAS.199..883B}.  The ratio of mass outflow to accretion rates is set to the standard value predicted by theory and numerical simulations of $0.3$.  The intial clump was taken to have a mass of $500 M_{\odot}$ and diameter of 1 pc., with a typical turbulent velocity field and observed mass to flux ratio.  

Figure 14 shows the results of these protostellar feedback simulations.  In the left panel without outflows, the SFE reaches nearly $49 \% $ with a total of 23 stars formed.  On the right, the simulation with outflows results in a SFE of $ 20 \% $ with 47 stars formed.  The results show that jets and outflows eject about a quarter of their parent molecular clump in high-speed jets. These extend out to distances of more than a parsec from their disks.  These outflows
reduce the star formation rate by about a factor of two, and  lead to the formation of $\sim 1.5 $ times as many stars compared to the no-outflow case. Perhaps the most important result is that MHD outflows reduce the average star mass by a factor of approximately three and may thus be essential for understanding the characteristic mass of the stellar initial mass function.   These results show that indeed, outflow feedback plays a significant role in reducing SFE in low mass clusters, bringing them down in agreement with the observations. 

\subsection{Planet formation and outflows} 

Given that star and planet formation both depend on how angular momentum in disks is disposed of, MHD wind torques are likely to play an equally profound role in the question of planet formation (see review \cite{2018haex.bookE.144P}).  There are at least three aspects of disk winds that may matter for planet formation and migration. 

The first is that braking and outflows lead to smaller disk radii, as noted in the observational section. This means that for a given disk mass, outflows produce disks with higher initial column densities.  This in turn implies that the column density of solids in the disk will be greater and this is likely to have an important effect on planetary accretion time scales and mass distributions (Alessi, Pudritz, and Cridland 2019, submitted).

Secondly, outflows may connect with the presence of rings and gaps in disks.  
Early arguments suggested that  gaps in the millimeter emission of disks could arise due to the enhanced growth of dust grains at opacity transitions (eg ice lines) at specific disk radii \citep{zhangbergin2015}.  Surveys now show, however, that there is little correlation between such opacity transitions and the positions of the gaps \citep{vandermarel2019}.  Much current work focuses on simulations that show that even a single, sufficiently massive planet can carve multiple gaps in inviscid ($\alpha \le 10^{-4}$) disks as a consequence of the spiral shock waves induced by planet-disk interaction (eg. \cite{dong2017}). This generally fits the ALMA observations of gaps and rings as well the origin of non-axisymmetric structures such as vortices, quite well \citep{zhangs2018}.  The formation of vortices requires low disk viscosity ($\alpha \simeq 10^{-4}$), which again suggests that low turbulence is necessary to explain key aspects of disk-planet interaction.  This picture does not address how such planets may have formed but does provide observational constraints on the masses and orbital radii of forming planets.  If disk winds drive disk accretion physics, then what specific effects might MHD disk winds have on creating these ring systems?   

Only a few papers have addressed this issue so far.  The launch of MHD (non-ideal limit) disk winds has been shown to also produce axisymmetric structures in the disks that could provide an explanation of the  ring systems \citep{2017A&A...600A..75B}.  In these simulations, zonal flows can be established in the disk that lead to the creation of density peaks that are anti-correlated with peaks in the vertical field. It has also been shown that in the outer parts of disks beyond 10 AU, where ambipolar diffusion dominates the diffusivity of the disk magnetic field, that strong sharply defined current layers can develop.  These drive fast flows that pinch the field, leading to their reconnection, which in turn leads slower gas accretion in these magnetically reduced regions.  Neighbouring regions have stronger fields and therefore undergo more rapid gas accretion - and it is here that gaps form \citep{suriano2018}.   This is demonstrated in both 2 and 3D non-ideal MHD simulations.  Gas in such regions is preferentially removed by the wind compared to the dust, enhancing the local dust to gas ratio perhaps to the point that streaming instabilities are triggered.  This would lead to rapid planetesimal formation, and perhaps planet formation.  It remains to be seen whether this model could predict planetary masses. 

Finally, the migration of planets in disks is sensitive to the mechanism of angular momentum transfer in disks. In MHD wind-driven regions  disks may be
considered to be inviscid, and there, corotation torques on the gas will arise from MHD disk
winds \citep{2017MNRAS.472.1565M, 2018MNRAS.477.4596M}. In this situation, the shape of the horseshoe orbit region very near the planet can be modified by the winds, leading to a
more history-dependent evolution of the horseshoe torque and its effects on planet migration. This may have important implications for how planets migrate when they are still low mass (so-called Type I migration). 

In summary, it looks as if disk winds are not only central to disk formation, evolution, and star formation - but they may also play a key role in planet formation.

\section{Conclusions}

The role of magnetic fields in outflows and star formation has undergone a rapid change in its perceived importance by the astronomical community over the last two to three years. The combination of ALMA observations, the spatial resolution of outflows, and theoretical and observational advances on the nature of MRI turbulence in disks all point towards MHD disk winds as a primary agent in these processes.  Our review then, comes to several important conclusions.

\begin{itemize}
    \item Magnetized outflow is likely the dominant mechanism of disk angular momentum transport. The basic predictions of MHD disk wind theory and a large, diverse body of numerical simulations are born out by ALMA observations. 
    \item The rotation of jets is now confirmed on several different scales.  This most basic prediction of MHD disk wind theory indicates that we are actually now observing how accretion disks dispose of their angular momentum.
    \item The observations show that a wide range of outflow velocities are observed.  The X-wind picture is "monochromatic" in that only a wind component originating at the magnetopause radius of the disk should be observed. While such a component may be present, it does not explain the wide range of data available.
    \item The observations support the universality of the MHD wind picture that is observed across all stellar masses.  The connection with jets in AGNs and micro quasars is also becoming much clearer. 
    \item Disk formation and observed disk properties arise from MHD turbulence conditions that dominate for the earliest pre-stellar states of forming cores. Even subsonic turbulence is sufficient to explain disk formation.  While hour glass magnetic geometries are seen in some systems, many more have disordered magnetic geometries indicative of these initial conditions.
    \item The feedback effects of magnetized outflows play a key role in regulating the star formation efficiency and masses of low mass stars and star clusters.  For massive star formation, the combination of MHD outflows and radiation pressure is central to massive star and massive cluster properties, but there are differences in the results.  More work on RMHD theory and simulations is needed.
    
\end{itemize}

The next five years of ALMA observations, more powerful numerical simulations, and advances in theory are likely to result in a paradigm shift from turbulent viscosity to MHD outflows as the fundamental basis of accretion disk theory and star formation.  This shift is already well underway.  One of the most exciting consequences of this sea change may well be in our understanding of planet formation.

% Please also refer to  \href{http://home.frontiersin.org/about/author-guidelines#Sections}{Author Guidelines} for further information on how to organize your manuscript in the required sections or their equivalents for your field

% For Original Research articles, please note that the Material and Methods section can be placed in any of the following ways: before Results, before Discussion or after Discussion.

\section*{Conflict of Interest Statement}

REP and TPR declare that there is no conflict of interest with any commercial or financial organization.

\section*{Author Contributions}

REP and TPR are both responsible for writing and editing the manuscript. 

\section*{Funding}
REP is supported by a Discovery Grant from the Natural Sciences and Engineering Research Council (NSERC) of Canada. TPR would like to acknowledge funding from the European Research Council under Advanced Grant No. 743029, Ejection, Accretion Structures in YSOs (EASY). 

\section*{Acknowledgments}
We thank the Editors for their invitation to write this timely review, and in particular Chris McKee for his patience and ongoing support of this work.  We are grateful to our referees, John Bally and Zi-Yun Li for their very useful and comprehensive reports.  We also thank Robi Banerjee for insightful discussions and commentary at various phases of this project.  Thanks also to Sylvie Cabrit, Gregory Lesur, Jake Simons, Aake Nordlund for lively discussions. 

\bibliographystyle{frontiersinSCNS_ENG_HUMS} % for Science, Engineering and Humanities and Social Sciences articles, for Humanities and Social Sciences articles please include page numbers in the in-text citations
\bibliography{bibreview}
%, bibreview2}

%%% Make sure to upload the bib file along with the tex file and PDF
%%% Please see the test.bib file for some examples of references

\end{document}